\documentclass[preprint,10pt]{aastex}

\usepackage[]{natbib}
\received{}   
\accepted{}     
\journalid{}{}
\articleid{}{} 

\begin{document}
\title{H$_{2}$ and CO emission from disks around T~Tauri and Herbig~Ae
pre-main-sequence stars and from debris disks around young stars: warm and cold circumstellar gas \altaffilmark{1}}

\author{W.F.\ Thi,\altaffilmark{2} E.F.\ van Dishoeck,\altaffilmark{2}
G.A.\ Blake,\altaffilmark{3} G.J.\ van Zadelhoff,\altaffilmark{2} J.\
Horn,\altaffilmark{4} E.E.\ Becklin,\altaffilmark{4} V.\
Mannings,\altaffilmark{5} A.I.\ Sargent,\altaffilmark{6} M.E.\ van den
Ancker,\altaffilmark{7} A.\ Natta,\altaffilmark{8}
J. Kessler\altaffilmark{3}}

\altaffiltext{1}{Based in part on observations with \textit{ISO}, an
ESA project with instruments funded by ESA Member States (especially
the PI countries : France, Germany, the Netherlands, and the United
Kingdom) and with participation of ISAS and NASA. }
\altaffiltext{2}{Leiden Observatory, P.O. Box 9513, 2300 Leiden, The
Netherlands.} \altaffiltext{3}{Division of Geological \& Planetary
Sciences, California Institute of Technology 150--21, Pasadena, CA
91125, USA.} \altaffiltext{4}{Department of Physics and Astronomy,
UCLA, Los Angeles, CA 90095--1562, USA.} \altaffiltext{5}{SIRTF
Science Center, MS 314-6, California Institute of Technology,
Pasadena, CA 91125, USA} \altaffiltext{6}{Division of Physics,
Mathematics and Astronomy, California Institute of Technology, MS
105-24, Pasadena, CA 91125, USA.} \altaffiltext{7}{
Harvard--Smithsonian Center for Astrophysics, 60 Garden Street, MS 42,
Cambridge, MA 02138, USA.} \altaffiltext{8}{Osservatorio Astrofisico
di Arcetri, Largo E. Fermi 5, I--50125 Firenze, Italy.}

\begin{abstract}
We present ISO Short-Wavelength-Spectrometer observations of H$_{2} $
pure-rotational line emission from the disks around low and intermediate
mass pre-main-sequence stars as well as from young stars thought to be 
surrounded by debris disks. The
pre-main-sequence sources have been selected to be isolated from molecular 
clouds and to have circumstellar disks revealed by millimeter interferometry.
We detect `warm' ($ T\approx 100-200 $~K) H$_{2} $ gas around
many sources, including tentatively the debris-disk objects. The mass of
this warm gas ranges from $ \sim 10^{-4} $ M$_{\odot } $ up
to $ 8\times  $10$^{-3} $ M$_{\odot } $, and can constitute
a non-negligible fraction of the total disk mass. Complementary single-dish
$ ^{12} $CO 3--2, $ ^{13} $CO 3--2 and $ ^{12} $CO 6--5 observations
have been obtained as well. These transitions probe cooler gas at
$ T\approx  $~20--80~K. Most objects show a double-peaked CO
emission profile characteristic of a disk in Keplerian rotation, consistent
with interferometer data on the lower-$ J $ lines. The ratios of
the $^{12} $CO 3--2/$^{13} $CO 3--2 integrated fluxes indicate
that $^{12} $CO 3--2 is optically thick but that $^{13}$CO
3--2 is optically thin or at most moderately thick. The $^{13} $CO
3--2 lines have been used to estimate the cold gas mass. If a H$_{2}$/CO
conversion factor of 1$ \times  $10$ ^{4} $ is adopted, the
derived cold gas masses are factors of 10--200 lower than those deduced
from 1.3 millimeter dust emission assuming a gas/dust ratio of 100,
in accordance with previous studies. These findings confirm that CO
is not a good tracer of the total gas content in disks since it can
be photodissociated in the outer layers and frozen onto grains in
the cold dense part of disks, but that it is a robust tracer of the
disk velocity field. In contrast, H$ _{2} $ can shield itself from
photodissociation even in low-mass `optically thin' debris disks and can
therefore survive longer. The warm gas is typically 1--10 \% of the total
mass deduced from millimeter continuum emission, but can increase
up to 100\% or more for the debris-disk objects. Thus, residual molecular
gas may persist into the debris-disk phase. No significant evolution
in the H$ _{2} $, CO or dust masses is found for stars with ages
in the range of 10$ ^{6} $--10$ ^{7} $ years, although a decrease
is found for the older debris-disk star $ \beta  $~Pictoris. The
large amount of warm gas derived from H$ _{2} $ raises the question
of the heating mechanism(s). Radiation from the central star  as well
as the general interstellar radiation field 
heat an extended surface layer of the disk, but
existing models fail to explain the amount of warm gas quantitatively.
The existence of a gap in the disk can increase the area
of material influenced by radiation. Prospects for future
observations with ground- and space-borne observations are discussed. 
\end{abstract}


\section{Introduction}

Recent discoveries of extra-solar giant planets stars have raised questions
about their formation \citep[e.g.][]{Butler99,MCM00}. Indeed, their
characteristics have been a surprise: they orbit much closer to the
stars than the planets in our own Solar System and their masses range
from that of Saturn up to 10 times the mass of Jupiter ($
M_{\rm{J}}\sim 10^{-3} $ M$ _{\odot } $). These planets are expected
to contain a solid core surrounded by a shell of metallic hydrogen and
helium and an outer low pressure atmosphere where hydrogen is in the
form of H$ _{2} $ \citep{Guillot99,CBLM00}.  To build such gaseous
giant planets, a large reservoir of H$ _{2} $ gas is needed at the
time of their formation, most likely in the form of a circumstellar
disk \citep[e.g.][]{BS96,Bod00}.

Most studies of circumstellar material associated with young stars
and debris-disk objects rely on continuum observations of the infrared
to millimeter emission produced by heated dust \citep[e.g.][]{BSCG90,SSB97}.
Dust particles represent only a trace component of disks, however,
which have 99\% of their mass initially in the form of H$ _{2} $
gas. Line imaging of trace molecules such as CO with millimeter interferometers
reveals the presence of gas in circumstellar disks with sizes of
$ \sim  $100-400~AU, but the inferred masses are up to two orders
of magnitude lower than those deduced from the dust continuum assuming
a standard gas/dust ratio and CO/H$ _{2} $ conversion factor as
in molecular clouds \citep[e.g.][]{KS95,MS97,Dut98,MS00,Dent95}. The millimeter
observations have nevertheless provided compelling evidence for gas
in Keplerian rotation around the central star \citep[e.g.][]{SDG00,Dut98}.
We report here the result of the first spectral survey of the pure-rotational
H${2} $ emission lines from circumstellar disks, the only molecule
hich can directly constrain the reservoir of warm molecular gas.

A related question is the temperature structure of the circumstellar
disks. The radial temperature structure is usually constrained by
modeling of the spectral energy distribution assuming either a thin,
flat disk geometry \citep[e.g.][]{ALS87} or a flaring disk
\citep[e.g.][]{KH87,C91}.  The dust in these models is heated by
radiation from the central star and by the release of energy through
accretion. Recent calculations by different groups show substantial
differences, however \citep[e.g.][]{Bell97,MHF99,DA98}.  Specifically,
flared disks may have surface layers with temperatures in excess of
100~K out to $\sim$100~AU \citep{CG97,CG99}.  The fitting of
spectral energy distributions is known to give ambiguous answers
and many disk parameters are still debated because of the
non-uniqueness of the fits \citep[e.g.][]{Hen98,BCILNS92}. 
H$_{2}$ emission line data can provide direct measurements of the
temperature of the warm gas.

According to standard models \citep[e.g.][]{Rud99,Lis93}, giant planet
formation by core accretion of gas occurs in the first few millions
years. Thus, the timescale for the disappearance of the gas compared
with that of the dust is of interest. Based on continuum data,
\citet{Strom89}, \citet{BSCG90}, \citet{OB95} and \citet{HaischLada01} 
suggested that dust
disks around T~Tauri stars disappear at an age of a few million
years. \citet{NGM00} searched for evolutionary trends in the outer
disk dust mass around Herbig~Ae stars. They found no evidence for
changes between 10$ ^{5} $ and 10$ ^{7} $ years, but an abrupt
transition seems to occur at 10$ ^{7} $ years from massive dust disks
to tenuous debris disks. \citet{Zuc95} conducted a survey of CO
emission from A-type stars with ages between 10$ ^{6} $--10$ ^{7} $
years and concluded that the gaseous disks disappear within 10$ ^{7} $
years. Determination of the gaseous mass from CO data is hampered,
however, by several difficulties compared to that from H$ _{2}
$. Provided that H$ _{2} $ traces the bulk of molecular gas, it can
constrain the time scale for gas dissipation from the disk directly.

Observations of the pure-rotational lines such as the H$ _{2} $
$ J $=2--0 S(0) 28.218 $ \mu  $m and $ J $=3--1 S(1) 17.035
$ \mu  $m lines are difficult from the ground because of the low
terrestrial atmospheric transmission in the mid-infrared. The Short
Wavelength Spectrometer (SWS) on board the {\it Infrared Space Observatory}
 (ISO) has allowed the first opportunity to observe a
sample of T~Tauri and Herbig~Ae stars, as well as a few young debris-disk
objects. The small mass of H$ _{2} $ implies that the two lowest
rotational lines have upper states which lie at rather high energies,
510~K and 1015~K above ground, respectively. The $ J $=2--0 and
$ J $=3--1 transitions are thus excellent tracers of the `warm'
($ T\approx  $ 80--200~K) component of disks. The mid-infrared H$_2$ data
provide complementary information to ultraviolet H$_2$ emission
\citep{VJL00} or absorption \citep{Roberge01} data toward
circumstellar disks, which either probe only a small fraction of the
H$_2$ or depend on the line of sight through the disk and foreground
material. H$_2$ has also been detected at near infrared wavelengths
\citep{WKB01}, but since these lines are excited by ultraviolet radiation, 
X-rays or shocks, they also cannot be used as a tracer of mass.

Spectroscopic observations of H$ _{2} $ have several advantages over
other indirect methods. First, since it is the most abundant gaseous
species, no conversion factor is needed. Also, contrary to CO, which
has a condensation temperature of $ \sim $20~K \citep{Yuri96}, it does
not freeze effectively onto grain surfaces unless the temperatures
fall below $ \sim $2~K \citep{SA93} --- lower than the minimum
temperature that a disk reaches. Its photophysics and high abundance
allow H$ _{2} $ to self-shield efficiently against photodissociation
by far-ultraviolet photons, such as those produced by A-type stars
\citep{Inga00}. Moreover, because the molecule is homonuclear, its
rotational transitions are electric quadrupole in nature, and thus
possess small Einstein A-coefficients. On the one hand, this presents
an observational problem since high spectral resolution is required to
see the weak line on top of the usually strong mid-infrared continuum.
On the other hand, the benefit is that the lines remain optically thin
to very high column densities, making the radiative transfer
simple. Another disadvantage is that the lines are only sensitive to
warm gas and cannot probe the bulk of the (usually) cold circumstellar
material probed by CO $ J=1-0 $ and $ J=2-1 $ interferometric
observations. Also, the high continuum optical depths at 28 $ \mu $m
prevent observations into the inner warm mid-plane of the disk. As a
complement, the same stellar sample has therefore been observed in the
$ ^{12} $CO and $ ^{13} $CO $ J=3-2 $ lines with the {\it James Clerk
Maxwell Telescope} and the $ ^{12} $CO $ J=6-5 $ line with the
\textit{Caltech Submillimeter Observatory}. These transitions probe
lower temperatures than H$_{2} $, about 20--80~K in the regime where
the dust is optically thin. The combination of H$ _{2} $ and CO
observations is sensitive to the full temperature range encountered in
disks. Along with millimeter continuum observations taken from the
literature, such data can provide a global picture of the structure
and evolution of both the gas and dust components of circumstellar
disks.

The paper is organized as followed. We first justify the choice of
the objects in our sample (\S 2). In \S 3, a description of the observations
is provided with emphasis on the special data reduction method used
for the H$ _{2} $ lines. In \S 4 and 5, the data are presented
and physical parameters such as mass and temperature are derived from
our observations of H$ _{2} $ and CO lines, as well as from 1.3
millimeter continuum fluxes taken from the literature. The accuracy
of each method is assessed. In \S 6, the different results are compared
and possible trends with effective temperature of the star or age
are investigated and the possible origin of the
warm gas is mentioned briefly. Finally, a discussion of the gas content 
in debris-disk objects is given.
The results for one object, the double binary GG~Tau,
have been presented by \cite{WF99a}. Earlier accounts of this work
may be found in \cite{vD98} and \cite{WF99b}, whereas the debris-disk
sources are discussed in \cite{WF01}. \cite{Stap99} present searches
for H$ _{2} $ emission in a complementary set of weak-line T~Tauri
objects.

\section{Objects}

Our study focuses on two classes of pre-main-sequence stars with
transitional ages spanning 10$ ^{6} $--10$ ^{7} $ years. T~Tauri stars
in the sample have spectral types of Me and Ke, corresponding to
stellar masses in the range from 0.25 to 2 M$ _{\odot } $ and are
probably younger analogs to the Sun. The higher-mass Herbig Ae stars
(2--3 M$ _{\odot } $) share the spectral type of debris-disk sources
and may be considered as younger counterparts to the debris-disk
objects.  In addition, three young debris-disk objects, namely 49~Ceti,
HD~135344 and $ \beta $~Pictoris are included in our sample. The
choice of objects is based on several criteria in order to maximize
the chance to detect the faint H$ _{2} $ lines on top of the
mid-infrared continuum and to avoid confusion with emission from
remnant molecular cloud material. First, the observed stars exhibit
the strongest 1.3 millimeter fluxes in the survey of T~Tauri stars by
\citet{BSCG90} and Herbig Ae stars by \citet{MS97,MS00}, i.e., they
possess the highest dust disk masses among the T~Tauri and Herbig Ae
stars in the Taurus-Auriga cloud. Second, they have all been imaged
with millimeter interferometers in CO and dust continuum and show
evidence for Keplerian disks. The only exceptions are UX Ori and WW
Vul, where no CO is detected. Third, the sample is biased toward
sources with a weak mid-infrared continuum at 10--30 $ \mu $m to
improve the line-to-continuum contrast.  This also prevents instrumental 
fringing problems. A faint mid-infrared excess suggests that a `dust hole'
exists in the disk close to the star, which may be caused by settling
and coagulation of dust particles in the mid-plane \citep{MN95}, to
clearing of the inner part of the disk by small stellar companion(s)
or proto-planet(s) \citep[e.g.][]{Lin00} or to shadowing of part of the 
disk \citep{Natta01}.  Finally, most of these
stars are located in parts of the Taurus cloud where the CO emission
is very faint or absent. Our original sample also included objects in
Ophiuchus \citep{vD98}, but these have been discarded from this sample
because of confusion by cloud material.

HD~135344, 49~Ceti and $ \beta  $~Pictoris have been identified
as debris-disk objects based on their far-infrared excess above the
expected photospheric flux level (e.g. Backman \& Paresce 1993). Keck
20 $ \mu  $m images reveal the presence of dust disks around the first two sources 
(Koerner 2000, Silverstone et al., private communication) whereas  $ \beta  $~Pictoris
has been imaged at many wavelengths \citep[e.g.][]{LBA01}. HD~135344, however, shows strong single-peaked 
H$ \alpha  $ emission \citep{DBR97}, suggesting that it also has Herbig Ae-type 
characteristics. The three debris-disk sources are objects located far from any 
molecular cloud.

This work does not constitute a statistical study since the sample
is limited in number and biased toward the highest disk masses and
low mid-infrared continuum. In Table~\ref{sources} the stellar
properties of objects of our sample are tabulated, including coordinates,
effective temperature, luminosity, and distance, together with references
to relevant literature.

\section{Observations}

\subsection{ISO-SWS observations}

The H$ _{2} $ $ J=2-0 $ S(0) line at 28.218 $ \mu  $m and
the $ J=3-1 $ S(1) line at 17.035 $ \mu  $m were observed with
the ISO-SWS grating mode AOT02 \citep{Gra96}. Typical integration
times were 600--1000~s per line, in which the 12 detectors were scanned
several times over the 28.05--28.40 and 16.96--17.11 $ \mu  $m
ranges around the lines. The H$ _{2} $ $ J=5-3 $ S(3) 9.66 $ \mu  $m
and $ J=7-5 $ S(5) 6.91 $ \mu  $m lines were measured in parallel
with the S(0) and S(1) lines, respectively, at virtually no extra
time. The spectral resolving power $ \lambda /\Delta \lambda  $
for point sources is $ \sim  $2000 at 28 $ \mu  $m and $ \sim  $2400
at 17 $ \mu  $m. The SWS aperture is $ 20''\times 27'' $ at
S(0), $ 14''\times 27'' $ at S(1), and $ 14''\times 20'' $ at
S(3) and S(5). For a few sources, observations of the S(1) line at
a $ 1' $ off position have been obtained as well. The S(2) J=4--2
12 $ \mu  $m line was also searched for toward 49~Ceti and HD~135344.

The continuum provides narrow band photometry. Since the observing
procedure does not perform spatial chopping, no zodiacal or background
emission is subtracted. The zodiacal background component has a continuous
spectrum corresponding to a dust temperature of about 260~K \citep{Reach96}
with an estimated flux density in the SWS aperture of about 0.3~Jy,
so that it can contaminate the continuum emission in some of our faintest
objects. Continuum fluxes above 3 Jy are considered as coming essentially
from the sources (star+disk) alone.

\subsubsection{Data reduction}

The expected peak flux levels of the H$ _{2} $ lines are close
to the sensitivity limit of the instrument. In order to extract the
H$ _{2} $ lines, special software designed to handle weak signals
on a weak continuum was used for the data reduction in combination
with the standard Interactive Analysis Package. The details and justification
of the methods used in the software are described elsewhere \citep{VT00}
and summarized below (see also the ISO-SWS manual at \url{http://www.iso.vilspa.esa.es$/$users$/$expl\_lib$/$SWS\_top.html}).

The raw data consist of 12 non-destructive measurements per elementary
integration (reset) corresponding to the 12 single-pixel detectors,
hence 24 observed points for a 2 second reset. A single scan lasts
200 seconds and typically 3--5 scans per line have been obtained,
corresponding to 7200--12000 data points. Since the readout system
acts as a capacitor, the signal has the form of an exponential decay,
and this curvature is first corrected using the AC time constant obtained
during the pre-flight calibration phase. Then a correction of the
instantaneous response function, or `pulse-shape', is applied with
the level of the correction determined from the data themselves because
the shape varies in time, a procedure called `self-calibration'. Finally,
a cross-talk correction is performed. This chain of calibration results
in removing the curvature and improving the straightness of the observed
slope which is in fact the measure of the flux. It also increases
appreciably the photometric accuracy and allows a better subsequent
determination of the noise.

Other factors, such as dark current drifts, influence the sensitivity
limit of the instrument as well and have to be corrected. The majority
of noisy data points are actually caused by impacts of cosmic rays,
called glitches, either on the detectors or on the readout electronics.
The level of cosmic ray hits fluctuates markedly, depending on the
position of the satellite and the activity of the Sun. The rate of
glitches may vary from scan to scan. At the level we are interested
in, up to 50\% of the data points can be rendered unusable by cosmic
rays or other instrumental artifacts.

Cosmic rays not only affect the sensitivity of the detectors instantaneously,
but also for some longer recovery time, a phenomenon called the post-glitch
effect. Most of the time, the glitches are secondary electron-hole
pairs created by the interaction of the energetic particles with the
detector elements; while the lifetime of these pairs is short, other
consequences of the impact can last longer. The decay of this effect
is observed to have an exponential form. The observing procedure used
by the SWS allows investigators to track events emerging simultaneously
in more than one detector. These so-called `correlated-noise events'
appear as a spurious feature in emission or sometimes in absorption
with a gaussian profile whose width is close to the resolution of
the instrument. The gaussian-like profile comes from the fact that
the glitch affects several detectors simultaneously, which results
in a shift in wavelength in the final spectrum.

In order to detect and circumvent the glitches, four types of statistical
filters have been defined. The first two are standard filters also
employed in the SWS pipeline software; the last two are additions
by us. The software is written in {\em IDL} (Interactive Data Language).
Each of these filters generates an array of non-valid points detected
by the adopted statistical method characterized by a unique parameter.
Thus, careful choices of filter parameters are crucial in determining
the quality of the resulting spectrum. The arrays are then cross-correlated.
Most of the time, the glitches are detected by more than one filter
and those points are immediately discarded.

The first filter consists in removing points which have a flux outside
a specified range defined by the user. This procedure may seem
artificial, but is justified by the fact that both line and continuum
fluxes are faint. In our data, this method removes points 5 sigma
above the continuum standard deviation calculated using all
points. The second filter searches for data points with a standard
deviation of the slope-fitting higher than the standard value adopted
in the SWS pipeline. This filter is efficient when used after the
self-calibration procedure described above. The third filter has been
set up specifically in this work to detect correlated noise. This
filter detects the glitches which are discrete stochastic events in
the time domain. The data from the 12 detectors observed at a single
time are summed, and the mean and standard deviations are computed. If
the standard deviation is higher than a specified parameter $ \phi $,
the data points are considered as glitches and are discarded in all 12
detectors. The value of $ \phi $, which is a multiple of the standard
deviation $ \sigma $, $ \phi =n\times \sigma $, is difficult to
determine {\it a priori} and can vary from scan to scan. Indeed, the
computed $ \sigma $ is affected by the number of cosmic ray hits --- a
high rate of glitches results in a high standard deviation--- so that
$ \phi $ has to be small. We have therefore used an automated
procedure to find the optimum values of $ \phi _{i} $, in which each
spectrum is examined with a range of values of $ \phi _{i} $ from $ n
$=1--6 times the standard deviation for each individual scan $ i $
with a step of 0.5. Thus, for a typical case of 3 scans, 10$ ^{3} $
versions of the reduced spectrum are generated. The fourth step
removes additional points one or two resets after a glitch is detected
by the previous technique. 

The data reduction procedure results in a
{}``dot cloud{}'' of observed fluxes as functions of wavelength. As a
final step, convolution with a gaussian whose FWHM is set by the
theoretical resolution of ISO-SWS at the relevant wavelength is
done. We have chosen to use a flux-conserving interpolation which can
modify the resolution but does not change the total integrated
flux. Since the lines are not spectrally resolved, the line profile is
not relevant. Small velocity shifts of the line of order $ 30 $ km s$
^{-1} $ compared with the rest wavelengths are frequent.  Many
parameters can cause such a shift, including the low signal-to-noise
of the data or pointing offsets. The latter problem not only affects
the peak position but also the flux since the beam profile is highly
dependent on the position in the entrance slit of the spectrometer.
Because the H$ _{2} $ emission can arise from a region 1--2$ '' $
offset compared to the position of the star, additional shifts of the
order of a few tens of km s$ ^{-1} $ are possible. The 1000 spectra
are then sorted by number of remaining data points. Generally, the
noise level due to glitches tends to decrease significantly as the number of
points decreases until a minimum is reached
when the statistical noise takes over
because of the small number of data points left. With this non-standard
data reduction procedure, it is difficult to devise an objective
detection criterion. Therefore, we adopt the following definition of the
level of confidence in our detections, depending on
the final $ S/N $ of the spectrum as well
as the fraction of reduced spectra in which the line is clearly
seen. A line is considered to be
detected when the $ S/N $ is 3 or higher and if its
profile lies within a gaussian mimicking the line profile of an
extended source filling the entire beam . Observations 
which are  only slightly
affected by cosmic ray hits show detections in a large number of the reduced
spectra ($ > $ 75\% of the 1000 spectra). The level of confidence of
the detection is considered ``high" in those cases.  The level becomes
``medium" when the detection is present in about 50--75\% of the
spectra. In cases of non-detection, the line is seen in less than 50\%
of the reductions.  Ultimately, we cannot rule out possible
instrumental artifacts which are not detected by our filters.

Of all the possible reductions, the spectrum with the lowest continuum
fluctuation (fringing) and noise and the highest $ S/N $ of the
line is kept as our best reduced spectrum and plotted in this paper.
The criterion of high peak flux and $ S/N $ comes from the fact
that the filters described above eliminate not only noisy data points
but also some valid points to a certain level. To keep this level
as low as possible, a compromise between quality (i.e., $ S/N $)
and flux level is adopted.

The non-gaussian nature of the noise makes the overall error difficult
to estimate, and we assume a fiducial 30\% photometric uncertainty in
the rest of the paper. This error is propagated into all the resulting
temperatures and masses. The actual uncertainty may be larger due to
the low $S/N$ of the data, but cannot be quantified in a consistent
way for different sources.  Note that the above procedure only throws
away data points and therefore cannot create artificial lines. This
is confirmed by the absence of lines at blank sky, or off-source,
positions reduced with the same procedure.

The above method was adopted for all sources with a weak continuum
level ($ < $3 Jy). For sources with a strong mid-infrared continuum
(AB Aur, HD 163296, RY Tau, CQ Tau, MWC 863), the fringing effect on
the continuum becomes the limiting factor for detection. Errors in the
dark current subtraction are a possible cause of this fringing. For
these sources, the fringes have been minimized by varying the dark
current level.

\subsection{CO observations}

As a complement to the ISO-SWS data, we have observed the same sample
of T~Tauri and Herbig Ae stars in various moderate- to high-$ J $ CO
transitions between 1998 and 2000 with submillimeter single-dish
telescopes. Previous studies have observed the lowest $ J $=1--0
and/or $ 2-1 $ transitions, either with interferometers
\citep{KS95,Dut96,MS97,MS00} or with single dishes
\citep[e.g.][]{Dut97}, but no homogeneous data set using the same
line, isotope and telescope exists for our sources.  We focus here on
the higher-$ J $ 3--2 and 6--5 transitions to probe gas with $
T $=20--80~K.

Observations of the $ ^{12} $CO and $ ^{13} $CO $ J=3-2 $ lines were
carried out at the {\it James Clerk Maxwell Telescope}
(JCMT)\footnotemark \footnotetext{The James Clerk Maxwell Telescope is
operated by the Joint Astronomy Centre on behalf of the United Kingdom
Particle Physics and Astronomy Research Council, the Netherlands
Organisation for Scientific Research, and the National Research
Council of Canada} using the dual polarization receiver B3 as the
frontend and the Digital Autocorrelator Spectrometer (DAS) as the
backend. Data were acquired with a beam switch of 180$ '' $ and, in
cases of extended emission, also a position switch up to $ 30' $. To
check for extended emission, several positions offset by $ 30''-60'' $
were observed as well.  Since the FWHM beam size of the JCMT at 345
GHz is 14$ '' $ and the extent of the disks at the distance of Taurus
is at most 5$ '' $, the observations suffer from large beam dilution,
as do the H$ _{2} $ data. The receiver was tuned single sideband, with
typical system temperatures above the atmosphere ranging from
400--600~K. The spectral resolution was typically 0.13 km s$ ^{-1} $,
sufficient to resolve the line profiles, but the data are
Hanning-smoothed once to improve the signal-to-noise. The final
spectral resolution is 0.26 km s$^{-1}$.  Integration times were
typically 10--20 minutes for $ ^{12} $CO and up to 2 hrs for $ ^{13}
$CO reaching a typical rms noise of $ \sim $ 15 mK. The antenna
temperatures have been converted to main beam temperatures using a
main beam efficiency at 330 GHz of $ \eta _{\rm{MB}} $=0.62 obtained
from observations of planets by the JCMT staff (see
\url{http://www.jach.hawaii.edu/JACpublic/JCMT/rx/b3/cal.html}).  The
data reduction was performed using the SPECX and CLASS software.

The $ ^{12} $CO $ J=6-5 $ data were obtained with the {\it Caltech
Submillimeter Observatory} (CSO)\footnotemark \footnotetext{The Caltech
Submillimeter Observatory is operated by the California Institute
of Technology under funding from the US National Science Foundation,
contract AST--9980846} using the sensitive 650 GHz receiver of \cite{Kooi98}
in double-sideband mode. Two acousto-optical spectrometers with resolutions
of 0.05 and 0.5 km s$ ^{-1} $ were used as the backends. Typical
system temperatures under excellent weather conditions were $\sim 2000$~K.
The CSO beam size at 650 GHz is $ \sim 14.5'' $, comparable to
the JCMT beam at 330 GHz, and the main beam efficiency is $ \eta _{\rm{MB}} $=0.40.

\section{H$ _{2} $ results and derived parameters}

\subsection{ISO-SWS spectra}

\begin{figure}
\figcaption[S0_aug00.eps]{The H$_2$ S(0) 28 $\mu$m
spectra observed with the ISO-SWS toward
pre-main-sequence and debris-disk stars. The underlying continuum has been
subtracted. The rest wavelength of the $J=$2--0 transition is indicated by the
dashed line. 
Small wavelength shifts may be attributed to instrumental effects (see text).
The dash-dotted gaussian
corresponds to emission by a source filling the beam; the H$_2$ lines have
to lie inside this gaussian to be considered detected.
 \label{h2s0}}
\plotone{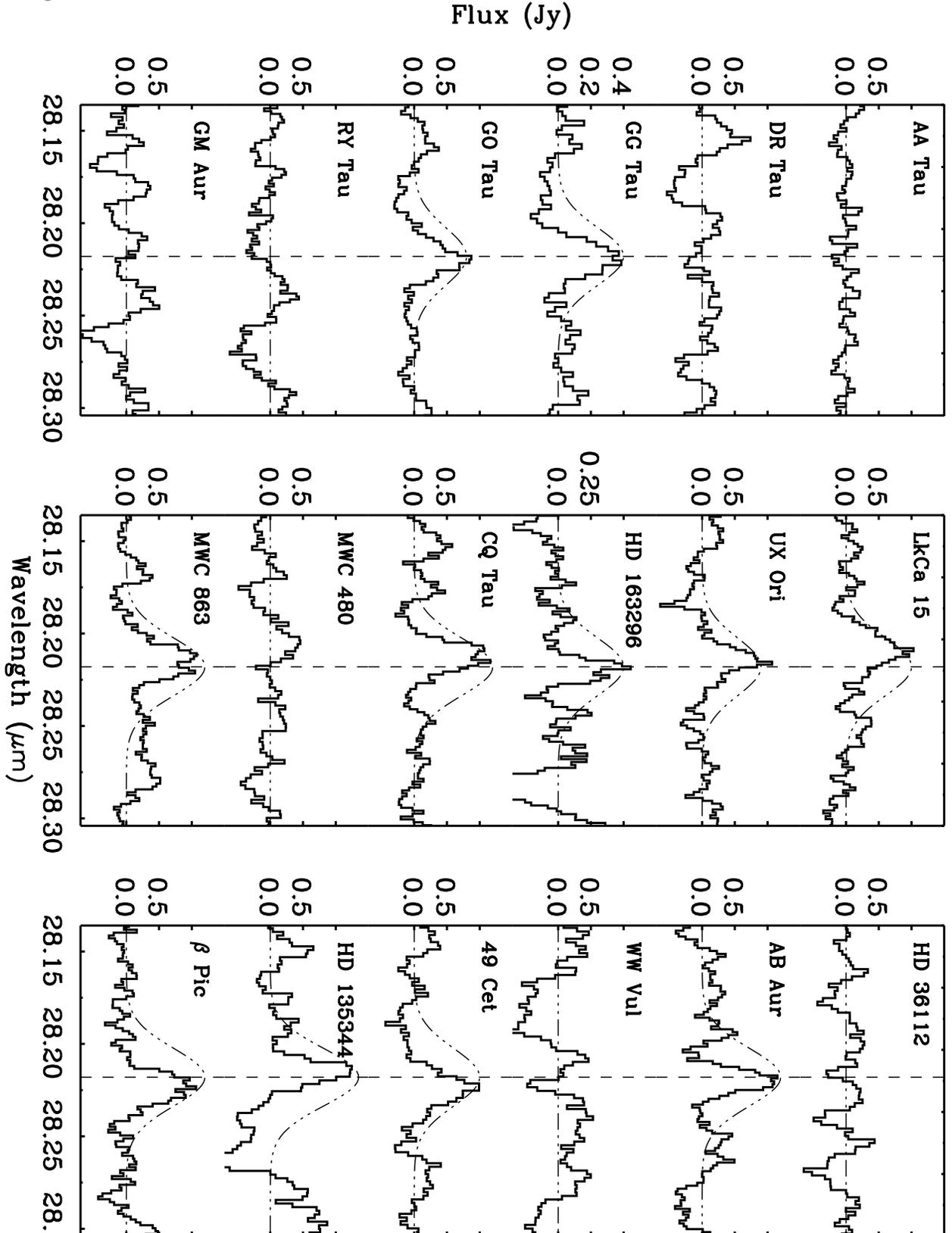}
\end{figure}

\begin{figure}
\figcaption[S1_aug00.eps]{As Figure~1, but for the 
H$_2$ S(1) J=3--1 tranistion. \label{h2s1}} 
\plotone{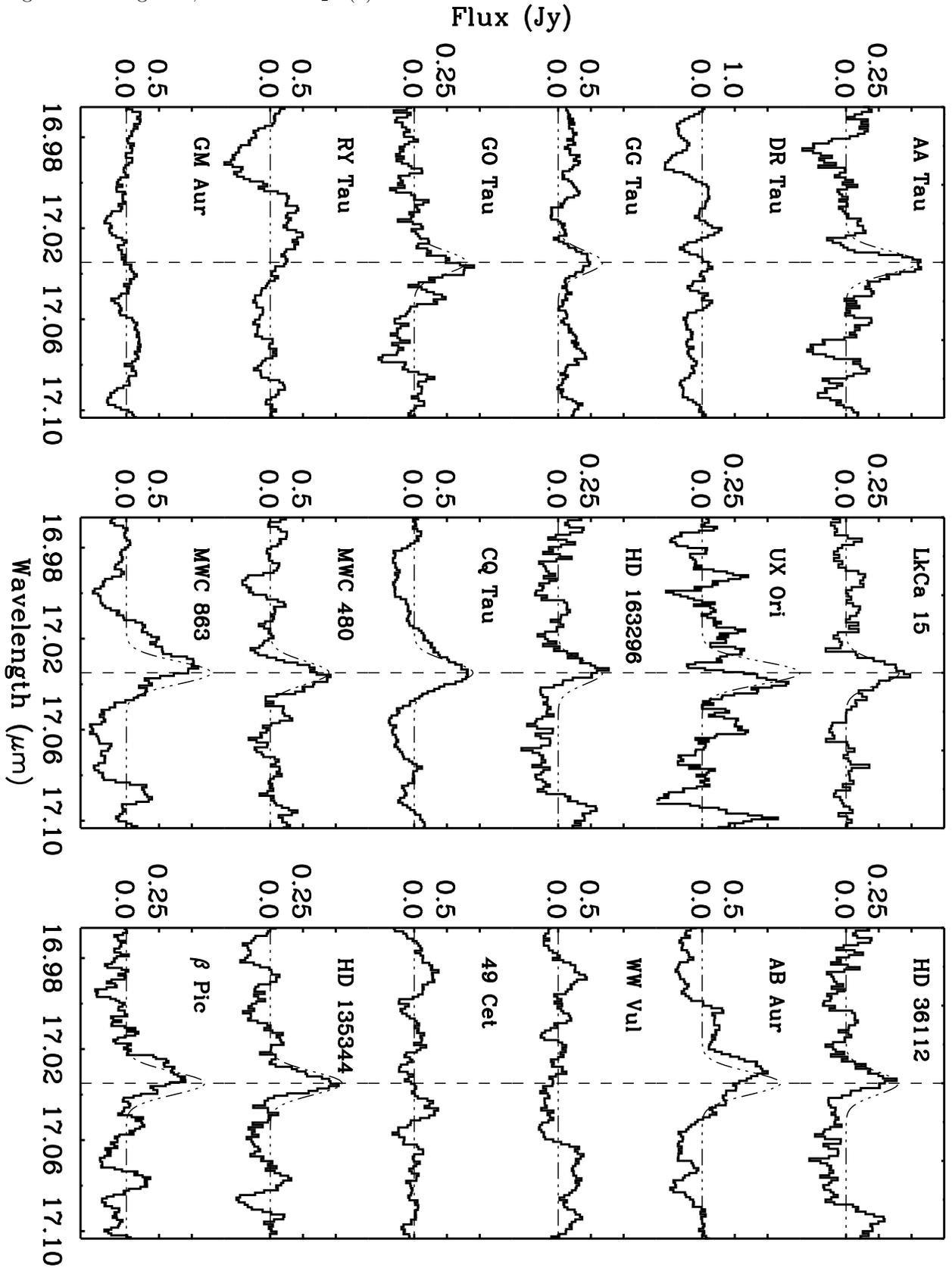}
\end{figure}

The final continuum subtracted H$ _{2} $ spectra are presented in
Figures~\ref{h2s0} and \ref{h2s1} for the $ J $=2--0 S(0) and
3--1 S(1) lines respectively. The typical rms noise level is 0.2--0.3
Jy. The dash-dotted lines in Figure~\ref{h2s0} and \ref{h2s1} indicate
the wavelength range in which the H$ _{2} $ line is expected, taking
into account the possible velocity shifts discussed in \S 3.  As
explained in \S 3, our line profiles may differ from the nominal
instrumental profile in width because of the adopted interpolation
scheme. The H$ _{2} $ line positions and basic molecular data are
listed in Table~\ref{h2data}, whereas the H$ _{2} $ S(0) and S(1) integrated
fluxes are reported in Table~\ref{h2iso}. The S(3) and S(5) lines are not
detected in any of the objects with an upper limit of $\sim $4$\times
$10$ ^{-15} $ erg s$ ^{-1} $ cm$ ^{-2} $ (3$\sigma $). The level of
confidence of a detection is indicated in the right-hand column of
Table~\ref{h2iso}. Similarly, the S(2) line is not detected toward 49~Ceti or
HD~135344 with an upper limit of $ \sim $9$\times $10$ ^{-15} $ erg s$
^{-1} $ cm$ ^{-2} $. Both the S(2) and S(3) lines are located in a
wavelength region where silicate features in emission or absorption
are strong.

Lines are detected in several disks around T~Tauri and Herbig Ae
stars, with no apparent trend with age or spectral type (see \S 6.2).
There are also likely detections of lines toward the debris-disk objects,
especially from HD~135344 and $\beta$ Pictoris.

The S(1) line shows a wider spread in observed fluxes and is more
readily detected for several reasons. First, the Einstein-$ A $
coefficient for the $ J $=$ 3-1 $ line is a factor of 16.5 larger than
that of the $ 2-0 $ line. Also, the spectral resolution is somewhat
higher at 17 $ \mu $m than at 28 $ \mu $m and the continuum lower, so
that the line-to-continuum ratio is larger. Finally, the sensitivity
of the 17 $ \mu $m detectors is better. All of these factors explain
why the S(1) line is more easily seen than the S(0) line, in spite of
the fact that the $ J $=3 level has a factor of 40 lower population
than the $ J $=2 level in gas with an estimated temperature of around
100~K.

\subsection{Contamination by diffuse H$ _{2}  $
emission?}

Except for the case of $ \beta $~Pictoris, the ISO-SWS beam is much
larger than the typical sizes of the circumstellar disks of $ < $5$ ''
$.  Thus, care has to be taken that the H$ _{2} $ emission is not
affected by any remnant cloud or envelope material in the
beam. Observations of the S(1) line have been obtained at several off
source positions $ 1' $ south. Toward 49~Ceti and HD~135344, which are
far away from any molecular cloud, no emission is detected off source
at the level of 8$ \times $10$ ^{-15} $ erg s$ ^{-1} $ cm$ ^{-2} $
rms, consistent with the expectation that diffuse atomic gas does not
emit in H$ _{2} $ lines. A weak S(1) line of $ \sim $ 10$ ^{-14} $ erg
s$ ^{-1} $ cm$ ^{-2} $ is seen $ 1' $ south of LkCa~15.  This flux
probably comes from a background cloud at a different velocity than
that of the source (see below).

Strong H$ _{2} $ lines have been detected with the SWS toward embedded
Herbig Ae and T Tauri stars where ultraviolet photons and shocks
interact with the surrounding material, but in these cases the
observed excitation temperatures of 500--700~K are much higher than
those found for our objects \citet{Mario00a,Mario00b}. Searches
for H$ _{2} $ lines toward diffuse molecular clouds with $ A_{V}=1-2 $
mag have been performed by \citep{WF99c}, but no lines are detected at
the level of 8$ \times $10$ ^{-15} $ erg s$ ^{-1} $ cm$ ^{-2} $ rms
for clouds with densities less than $ 10^{3} $ cm$ ^{-3} $ and
incident radiation fields less than 30 times the standard interstellar
radiation field. The strengths of the S(0) and S(1) lines from diffuse
clouds can also be estimated from ultraviolet observations of H$ _{2}
$ obtained with the {\it Copernicus} satellite and the
\textit{Far-Ultraviolet Space Explorer} (FUSE) \citep{SJ75,Shull00}.
Consider as an example the recent FUSE results for the translucent
cloud toward HD 73882 ($ A_{V} $=2.4 mag) by \citet{Snow00}. The
observed column densities in $ J $=2, 3 and 5 translate into fluxes of
$ 2.7\times 10^{-14} $, $ 1.2\times 10^{-13} $ and $ 1.1\times
10^{-13} $ erg s$ ^{-1} $ cm$ ^{-2} $ for the S(0), S(1) and S(3)
lines, respectively, assuming the gas fills the ISO-SWS beam. The S(0)
and S(1) fluxes are comparable to our observed values, but the S(3)
flux is significantly higher than our upper limits. Indeed, both the
{\it Copernicus} and FUSE data give typical excitation temperatures
for the $ J $=2--7 levels of $ \sim $300~K, significantly larger than
the values of $ \sim $100--200~K found here. Moreover, such thick
clouds as those toward HD~73882 or $ \zeta $~Oph emit significant CO
emission \citep[e.g.][]{GvDB94,vD91}, which is generally not observed
at the off source positions in our sample.

To check for the presence of molecular gas at off-source positions,
mini-maps in $ ^{12} $CO 3-2 have been obtained in steps of 30$ '' $
for most of our sources (see also \S 5). For GG~Tau, LkCa~15, MWC~480
and GM~Aur, no emission is found off source down to 30~mK rms at the
velocity of the sources. In other cases such as GO~Tau, DR~Tau,
HD~163296, weak off-source emission is seen in $ ^{12} $CO 3--2
with velocities shifted compared to the source velocity, but
this off-source emission is
not seen in $ ^{13} $CO 3--2. The off-source CO emission is at least
a factor of 30 lower than found for translucent clouds such as HD~73882.
Thus, the bulk of the molecular gas for these sources is clearly located
in the disks, but some lower-density cloud material may be present.
Based on the above arguments combined with the absence of S(3) emission,
this diffuse gas is expected to make only a small contribution to
the S(0) and S(1) lines. Finally, two of our objects, AB~Aur and
RY~Tau, show single dish CO data which are clearly dominated by more
extended remnant envelope material. In these cases, a significant
fraction of the H$ _{2} $ emission may arise from extended gas
although the temperature in the envelope (10--20 K) may be too low
to produce substantial rotational excitation.

In summary, for most of our sources, the H$ _{2} $ emission is
unlikely to be contaminated by extended emission from diffuse
molecular gas, but this cannot be ruled out for cases such as AB
Aur. In fact, H$_2$ ultraviolet absorption toward AB Aur has been
detected by FUSE \citep{Roberge01} and arises in an extended
low-density envelope around the star or from general foreground
material.

\subsection{H$ _{2} $ temperatures}

The integrated flux $ F_{ul} $ of a rotational emission line $
J_{u}\rightarrow J_{l} $ of H$ _{2} $, assuming that the line is
optically thin and not affected by dust extinction, and that the gas
is at a single temperature $ T_{\rm{ex}} $, is given by
\begin{equation} 
F_{ul}=\frac{hc}{4\pi \lambda
}N({\rm{H}_{2}})A_{ul}x_{u}\Omega \, \, \, \rm{erg}\, \, \rm{s}^{-1}\, \,
\rm{cm}^{-2},\rm{ }
\end{equation}
where $ \lambda  $ is the wavelength of the transition, $ A_{ul} $
is the spontaneous transition probability, $ N $(H$ _{2} $)
the total column density of H$ _{2} $ and $ x_{u} $ the population
of level $ u $. $ \Omega  $ corresponds to the source size,
which is not known since the H$ _{2} $ data are spatially unresolved.
For gas densities larger than 10$ ^{3} $ cm$ ^{-3} $, the lines
are thermalized and the population $ x_{u} $ follows the Boltzmann
law \begin{equation}
x_{u}=(2J_{u}+1)g_{N}\exp {\left( -E_{J_{u}}/kT_{\rm{ex}}\right) }/Q_{\rm{H}_{2}}(T_{\rm{ex}}),
\end{equation}
 with $ E_{J_{u}} $ being the energy of the upper level, $ Q_{\rm{H}_{2}}(T_{\rm{ex}}) $
the partition function of H$ _{2} $ at $ T_{\rm{ex}} $ and
$ g{_{N}} $ the nuclear statistical weight factor, which is 1 for
para-H$ _{2} $ (even $ J $) and 3 for ortho-H$ _{2} $ (odd
$ J $). The lines are optically thin up to column densities of
$ 10^{23} $ cm$ ^{-2} $ owing to the low values of the Einstein
$ A- $coefficients.

When both the S(0) and S(1) lines are detected, the excitation temperature
can be obtained from the relation \begin{equation}
T_{\rm{ex}}=\frac{505.24}{\ln \left( 112.51\times F_{20}/F_{31}\right) }\, \, {\rm{K}.}
\end{equation}
 In LTE, $ T_{\rm{ex}} $ is equal to the kinetic temperature
$ T $. $ F_{20} $ and $ F_{31} $ are the integrated S(0)
$ J=2-0 $ and S(1) $ J=3-1 $ fluxes, respectively. Since no
data are available to constrain the ortho-H$ _{2} $ to para-H$ _{2} $
ratio, we assume that the ortho/para ratio is in LTE at the temperature
$ T=T_{\rm{ex}} $. At $ T $=100~K, the ortho/para ratio
is 1.6. No correction for differential extinction between the S(0)
and S(1) lines is applied.

The inferred temperatures range from 100 to 200~K (see Table~\ref{h2iso}).
The uncertainty of 30\% in the fluxes propagates into a $ \sim  $10\%
error in the temperature. If the emission were affected by $ \sim  $30
magnitudes of extinction, the derived temperatures would be increased
by typically $ \sim  $20~K, illustrating that this does not have
a large effect. The upper limits on the S(3) line translate into upper
limits on the gas temperature of typically $ < $~250~K if no
correction for differential extinction is made. Similarly, the upper
limits on the S(2) line imply temperatures $ < $200~K for HD 135344
and 49~Ceti. The detection of either the S(0) or S(1) lines combined
with the upper limits on S(3) imply a probable temperature range of
100--200~K for the gas.

\subsection{Warm gas masses from H$ _{2}  $}

\begin{figure}
\figcaption{Mass of H$_2$ as a function of the excitation temperature 
for an observed H$_2$ S(1) integrated flux of 10$^{-13}$ 
erg s$^{-1}$ cm$^{-2}$ for a source at 140 pc. 
The dash-dotted line corresponds to the 
typical Solar Nebula mass (10$^{-2}$ M$_{\odot}$) and the dashed line 
indicates the mass of Jupiter. \label{masstemp}}
\plotone{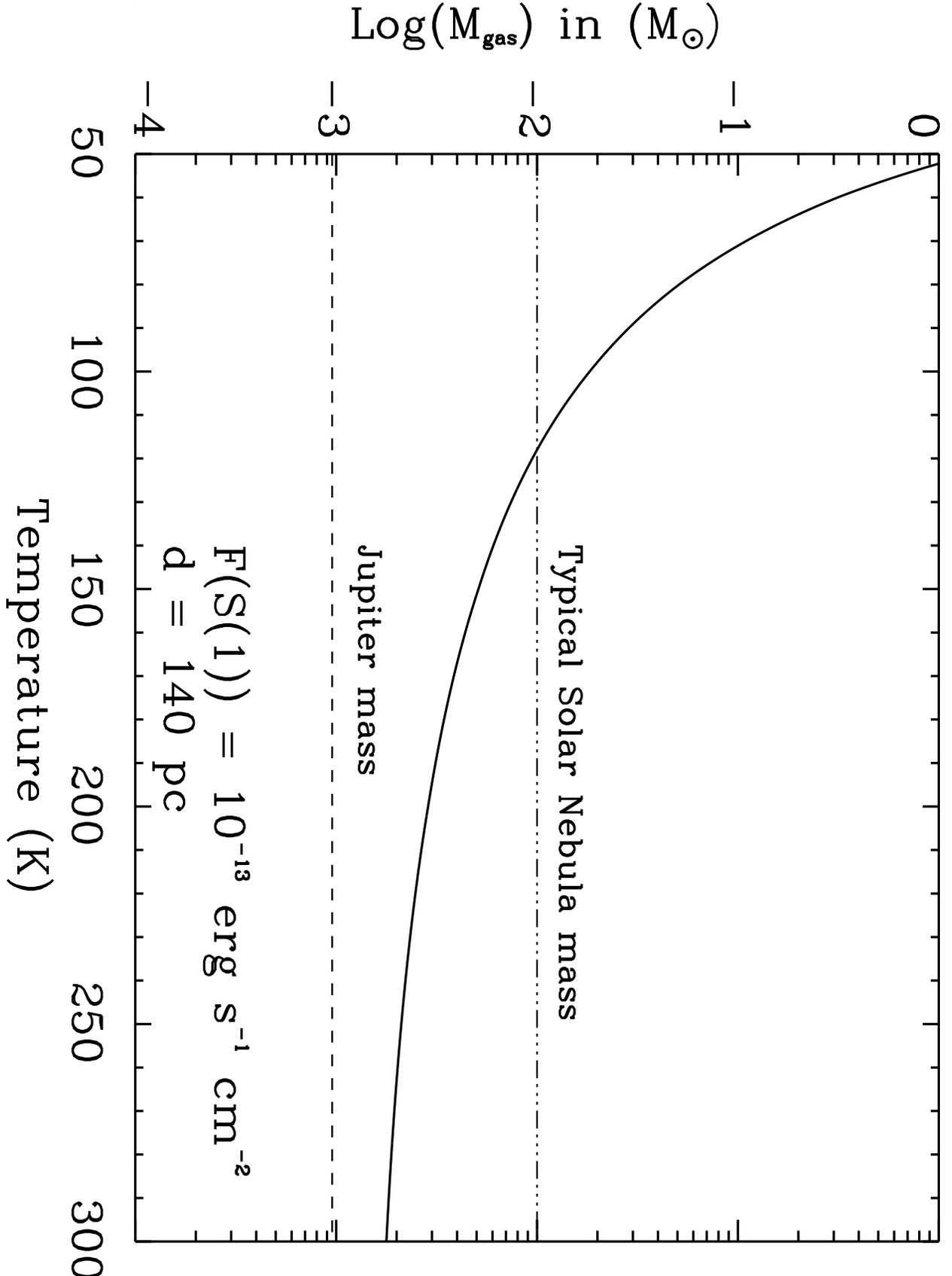}
\end{figure}

Because the lines are optically thin, the measured fluxes can be translated
directly into a beam-averaged column density of warm gas using Eq.\ (1)
with $ \Omega  $ equal to the solid angle of the ISO-SWS beam at
the observed wavelength. The gas mass can be computed from \begin{equation}
M_{\rm{warm}\, \, \rm{gas}}=1.76\, \, 10^{-20}\, \, \frac{F_{ul}\, \, d^{2}}{\left( hc/4\pi \lambda \right) A_{ul}x_{u}}\, \, {\rm{M}_{\odot }}
\end{equation}
 where in addition to the above assumptions, $ d $ is the distance
in pc which is provided by the Hipparcos satellite or taken from the
literature (see Table~\ref{sources}). The derived masses are presented in Table~\ref{h2iso}
and depend strongly on the population $ x_{u} $ and thus on the
temperature $ T $. This is illustrated in Figure~\ref{masstemp},
which shows the inferred mass as a function of temperature for a S(1)
line flux of 10$ ^{-13} $ erg s$ ^{-1} $~cm$ ^{-2} $ at
the distance of Taurus (140 pc): for temperatures between 100 and
200~K, the mass changes by approximately one order of magnitude.
Assuming an error on the flux of $ \sim  $30\% and including an
error on the distance of 10\%, the error on the mass reaches $ \sim  $55\%
in cases where both the S(0) and S(1) lines have been measured. When
only one line is detected, a range of masses is obtained by assuming
that the excitation temperature lies between 100 and 200~K and consequently
shows a large spread.

If the line emission were affected by 30 magnitudes of extinction,
the derived masses are changed by less than 20\%: the increase in
mass due to the extinction correction is compensated by its decrease
owing to the higher inferred temperature (see \S 4.3).

\subsection{Mid-infrared continuum: tracing the warm dust}

The continuum around the lines can be used to perform narrow band
photometry and the resulting absolute fluxes are given in Table~\ref{mircont}.
Moreover, data in the 3.4 $ \mu  $m region have been obtained in
parallel and are included. The values at 28 $ \mu  $m
are consistent, within the errors (estimated to be $\sim$ 30\%), with the IRAS Point 
Source Catalog
fluxes extrapolated from observations at 25 $ \mu  $m. As mentioned
in \S 3, the sources with faint mid-infrared continuum fluxes ($ < $~1~Jy)
can be contaminated by zodiacal emission and include a contribution
from the star. Typically, for a star located in the Taurus cloud with
an effective temperature $ T_{\rm{eff}} $=8700~K and luminosity
$ \log (L/L_{\odot }) $=1.5, the stellar monochromatic fluxes are
0.26, 0.04, 0.16, 3 10$ ^{-3} $ and 7 10$ ^{-4} $ Jy at 3.4,
6.9, 9.6, 17 and 28 $ \mu  $m respectively. Thus, the stellar contributions
at 17 and 28 $ \mu  $m are negligible compared to the zodiacal
emission.

A complete understanding of the mid-infrared line and continuum
emission requires a detailed radiative transfer code and a specific
disk model implying many assumptions. We adopt here a simplified
picture based on the \cite{CG97} model of irradiated passive disks.
The disk is divided into 3 components: (i) a hot part giving rise to
the near-infrared emission; (ii) a warm part ($ T_{\rm{dust}}\approx$80--300~K) 
responsible for the mid-infrared emission, and perhaps also
the H$ _{2} $ emission; and (iii) a cold part ($ T_{\rm{dust}}< $80~K)
giving the submillimeter continuum and the CO emission. Component (ii)
corresponds to the warm surface layer in the Chiang \& Goldreich
models. Component (i) is not present in those models, but may be due
to very hot thermal emission in an inner boundary layer,
or to non-thermal emission by very small grains or PAHs
\citep{Mario00b,SSB97}, from Fe-containing grains
\citep{BKAW00} or to due a very hot inner layer \citep{Natta01}.
Our main reason for this partition is to compare
separately the `warm' and `cold' gas and dust components.

\begin{figure}
\figcaption{Dust temperature as a function of the ratio of the continuum 
fluxes at 17$\mu$m and at 28$\mu$m, assuming optically thin emission.
The vertical bars at the bottom indicate the observed ratios for
our sources. The errors on the ratio are typically 50\%. \label{figcont}}
\plotone{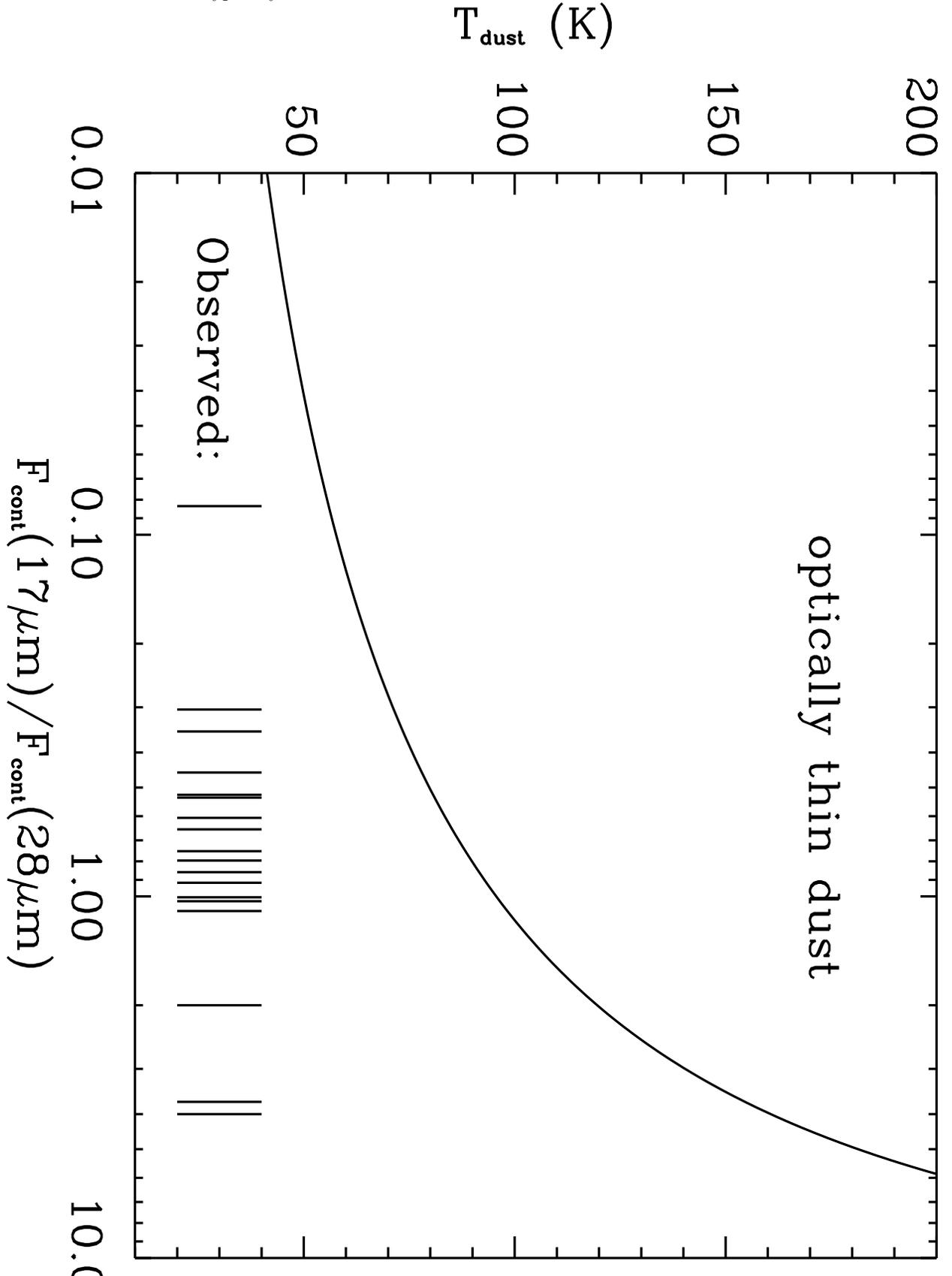}
\end{figure}

\begin{figure}
\figcaption{Warm gas temperatures versus warm dust temperature for sources having both S(0) and S(1) lines detected. The errors on the temperatures are $\sim$10\%. \label{Td_g}}
\plotone{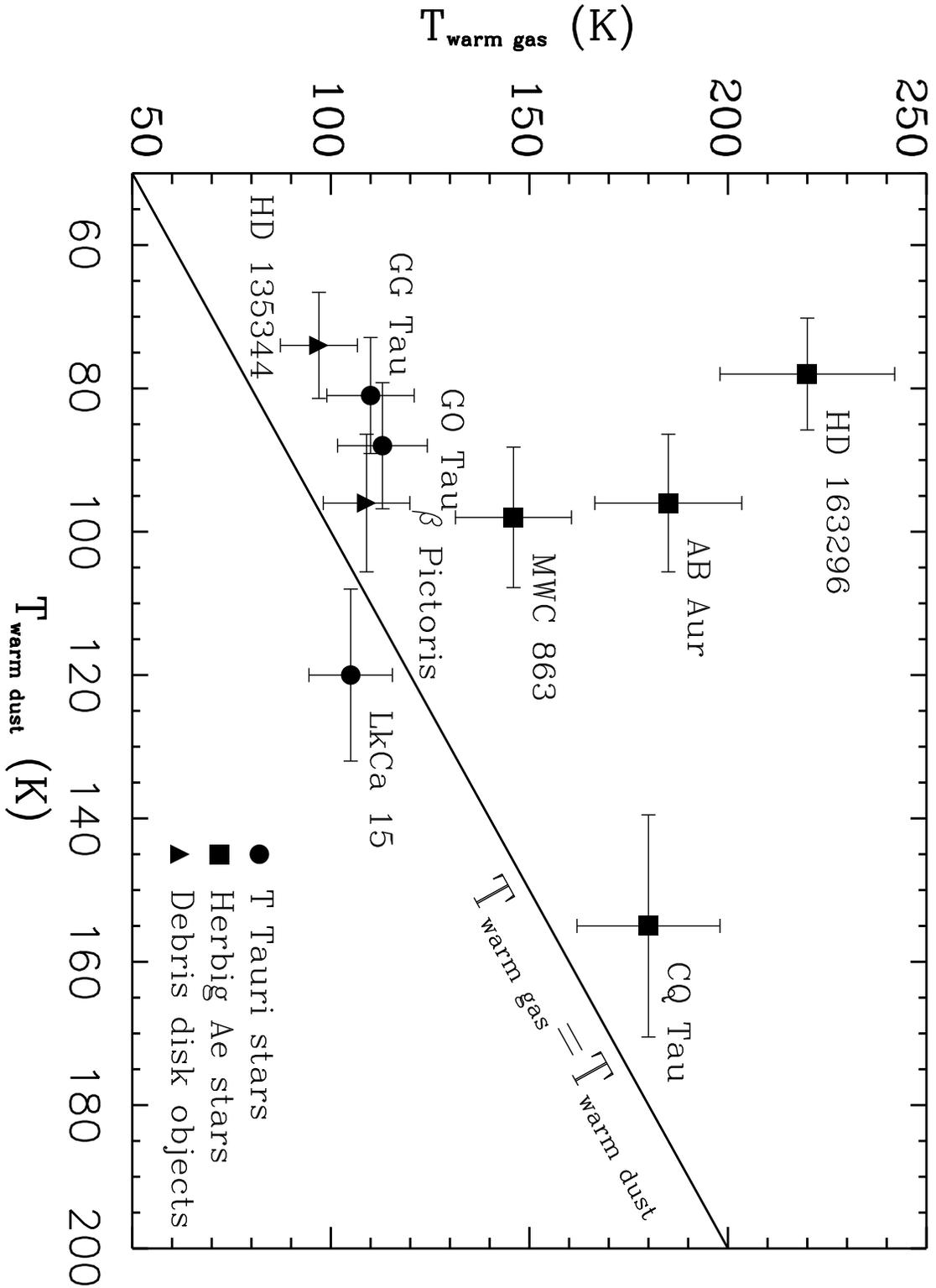}
\end{figure}

To obtain a rough estimate of the temperature of the warm dust, the
17/28 $\mu$m flux ratios have been fitted with an optically thin dust
model, as may be appropriate for the surface layers of disks.  The
grain emissivities of \citet{OH94} have been
used. Figure~\ref{figcont} shows the resulting warm dust temperature
for different values of the 17/28 $ \mu $m ratio. The observed values
are included in Figure~\ref{figcont} and the resulting fits are
summarized in Table~\ref{mircont}.  The observational errors on the
temperature are $\sim$10\%. Interestingly, the beam average warm gas
temperatures derived from H$ _{2} $ are higher by 20--50~K (see
Figure~\ref{Td_g}). There are several possible explanations of this
difference, e.g.\ the location of the emitting gas and dust may be
different or a gas heating mechanism other than gas-grain collisions
has to be invoked.

\section{CO results and derived parameters}

\subsection{$ ^{12}  $CO and $ ^{13}  $CO
3--2 lines}

\begin{figure}
\figcaption{Single-dish JCMT spectra of the $^{12}$CO 3--2 line at 345 GHz 
toward 
protoplanetary disks. The vertical scale is $T_{\rm MB}$ in K. The horizontal 
scale denotes the LSR velocity. The characteristic double-peaked
line profile due to a rotating disk is seen for many sources. \label{12co}}
\plotone{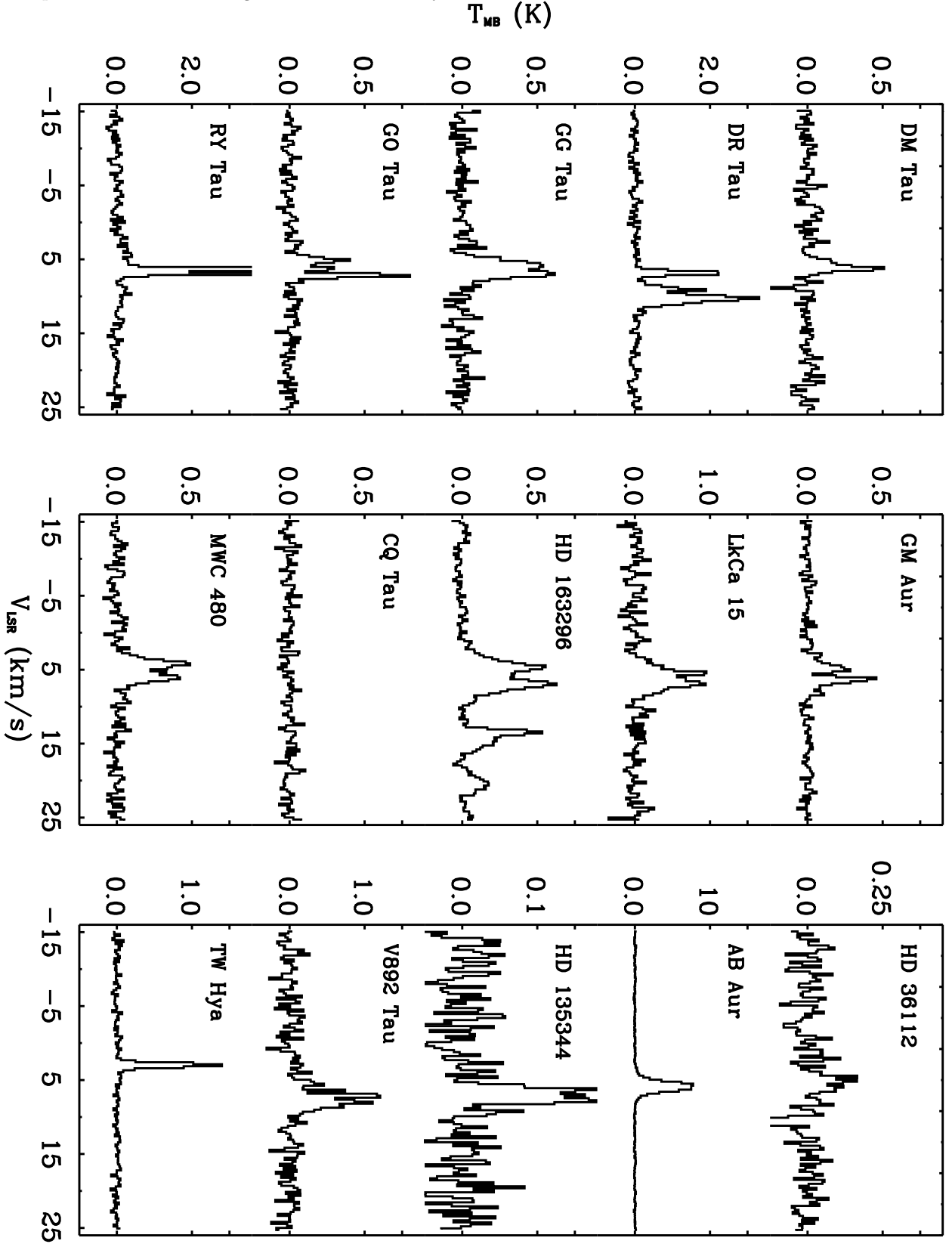}
\end{figure}

\begin{figure}
\figcaption{As Figure~6, but for $^{13}$CO 3--2 JCMT spectra. \label{13co}}
\plotone{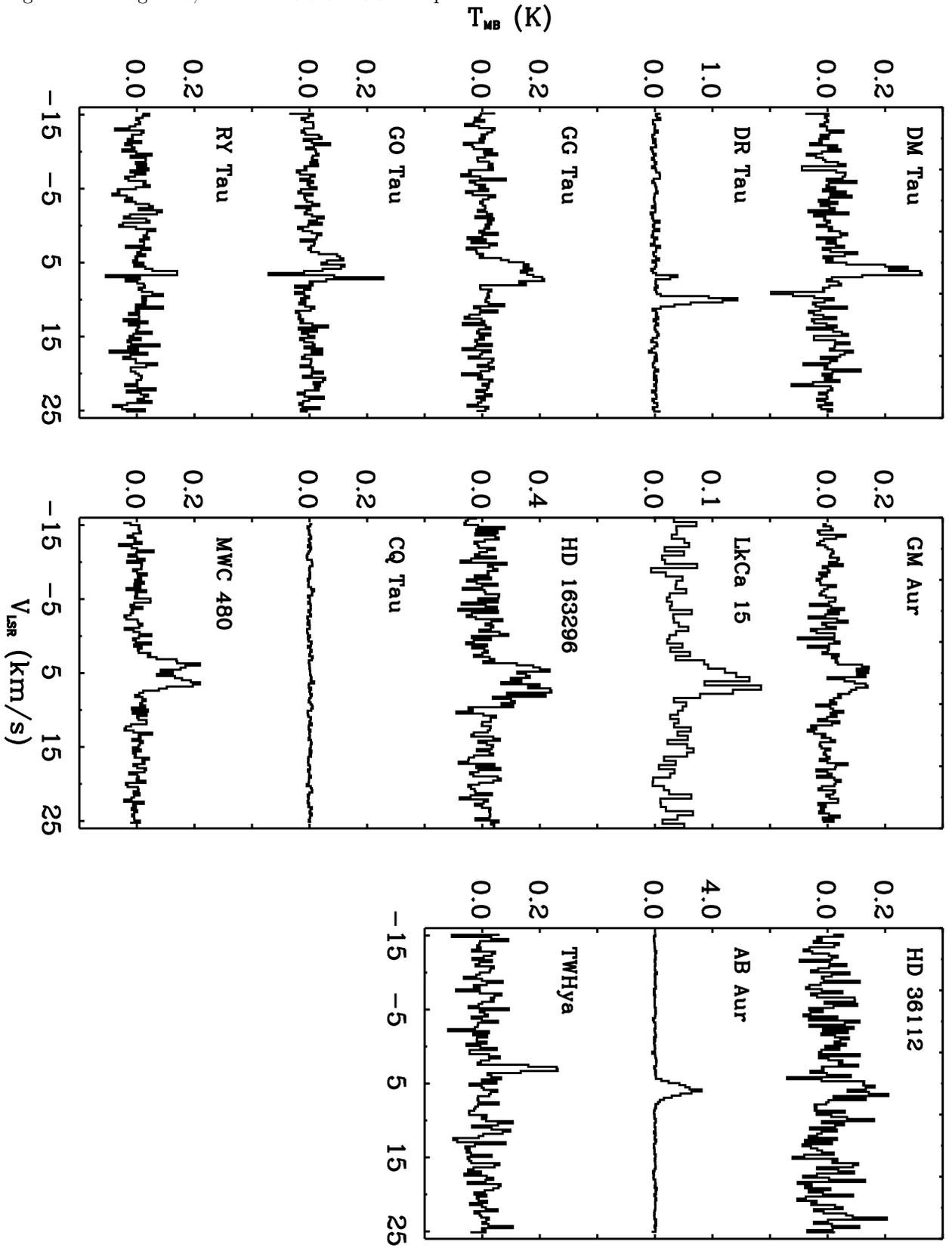}
\end{figure}

Observations of $ ^{12} $CO 3--2 lines have been performed with the
JCMT toward most of the sources observed with ISO, plus a few other
T~Tauri and Herbig~Ae stars. In all but a few cases, the $ ^{12} $CO
3--2 and $ ^{13} $CO 3--2 lines are detected with good signal-to-noise
($>$ 5 $\sigma$), and the spectra are presented in Figures~\ref{12co}
and \ref{13co}.  The lines show the typical double-peaked profile
consistent with emission from a disk in Keplerian rotation seen at a
certain angle \citep[e.g.][]{BS93,Gui99}.  The full width at half
maximum of the line profile is typically 2.5--3 km s$ ^{-1} $ and the
separation between the two peaks is of order of 1.2--2 km s$ ^{-1}
$. The integrated fluxes are computed by fitting two gaussians, which
are tabulated in Table~\ref{co3_2}. The uncertainty in the integrated fluxes is
dominated by the calibration error of $ \sim $30 \%. The mean
integrated area of the $ ^{12} $CO 3--2 line for the T~Tauri stars is
$ \sim $0.5 K km s$ ^{-1} $ higher than that for Herbig~Ae stars. The
clear presence of the double peak suggests that the microturbulence in
these disks is no more than 0.2-0.3 km s$ ^{-1} $, comparable to the
thermal width of $ \sim $0.22 km s$ ^{-1} $ at 30~K. CO 3--2 is not
detected toward CQ~Tau. This non-detection is compatible with the 
CO 2--1 flux detected by \cite{MS00} using the Owens Valley Millimeter Array.

A clear $ ^{12} $CO 3--2 disk line profile is also detected toward the
debris-disk object HD 135344, previously studied in $ ^{12} $CO 2--1 by
\cite{CW95}. No 3--2 searches have been performed toward 49~Ceti, but
Zuckerman et al.\ (1995) report a detection of the CO 2--1 line. A
deep search for $ ^{12} $CO 2--1 has been performed toward $ \beta
$~Pictoris by \cite{LA98} with a limit of 11 mK rms in the 23$ '' $
SEST beam. CO is seen by ultraviolet absorption lines, however, and
\cite{Roberge00} infer a column density of (6.3~$ \pm $0.3) 10$ ^{14}
$ cm$ ^{-2} $ of CO gas at a temperature of 20--50~K.

\begin{figure}
\figcaption{Velocity integrated $^{12}$CO 3--2 flux in K km s$^{-1}$ plotted 
against velocity integrated $^{13}$CO 3--2 flux normalized at 
100 pc.\label{coratio} }
\plotone{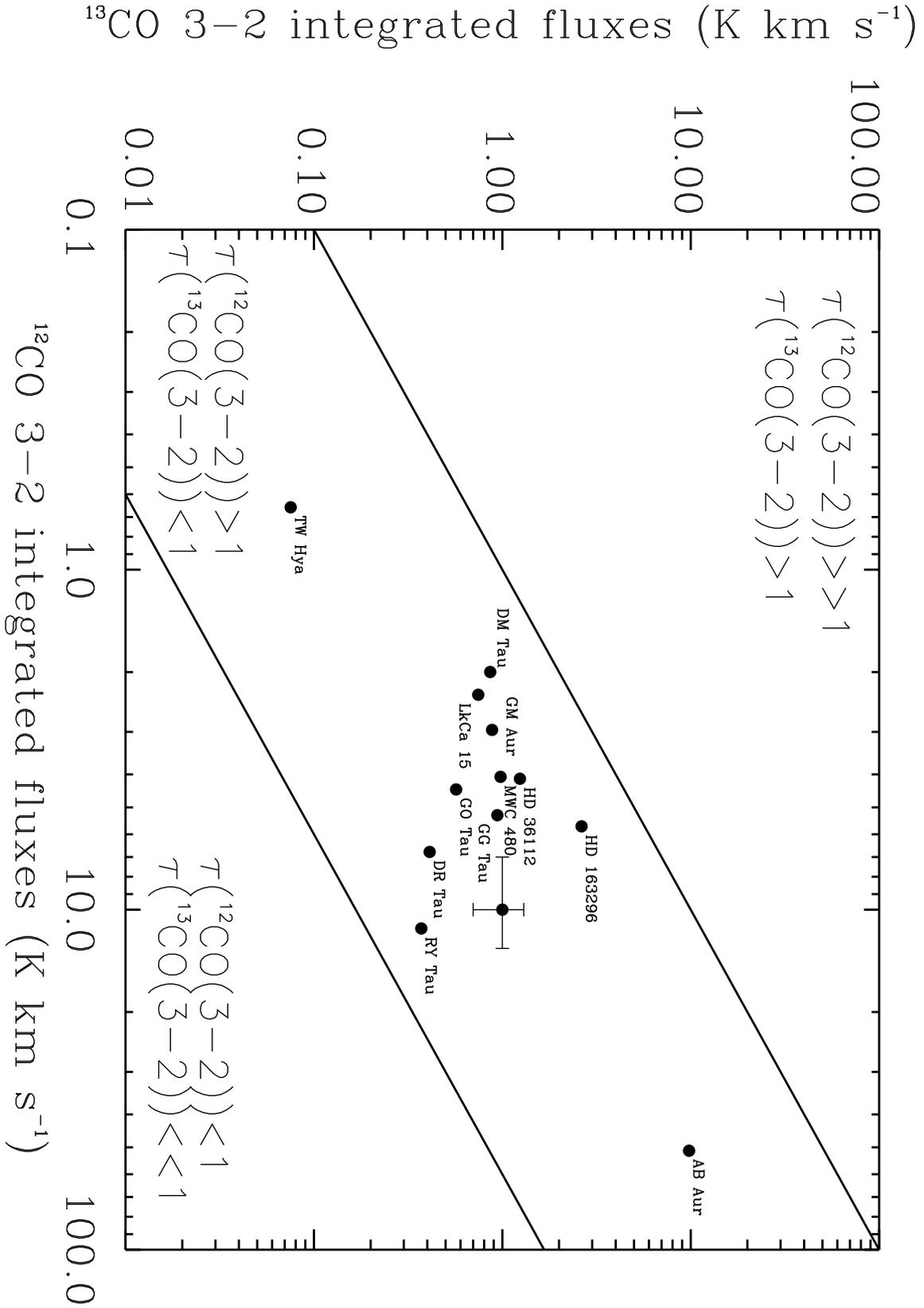}
\end{figure}

Figure~\ref{coratio} shows the $ ^{13} $CO 3--2 versus $ ^{12} $CO
3--2 integrated line fluxes normalized to a distance of 100~pc. The
different regimes of optical depth are indicated. The data fall in
the region where $ ^{13} $CO 3--2 is thin whereas $ ^{12} $CO
3--2 is optically thick. No difference is found between the T~Tauri
and the Herbig~Ae stars. Assuming a [$ ^{12} $C]/[$ ^{13} $C]
ratio of 60 and the same excitation temperature for $ ^{12} $CO
and $ ^{13} $CO, the beam-averaged optical depths $ \bar{\tau } $
of the $ ^{13} $CO 3--2 line can be calculated, and are given in
the last column of Table~\ref{co3_2}. If the excitation temperature of $ ^{12} $CO
is higher than that of $ ^{13} $CO, as suggested by models of \citet{GJ01},
$ \bar{\tau } $ could be increased by a factor of two. Nevertheless,
a low optical depth $ \bar{\tau }<1 $ of $ ^{13} $CO 3--2 is
confirmed by the non-detection of C$ ^{18} $O 3--2 emission \citep{GJ01}.
Thus, the $ ^{13} $CO 3--2 line could constitute a tracer of the
total gas mass in the outer part of disks provided the excitation
temperature can be determined and the $ ^{13} $CO/H$ _{2} $
conversion factor is known.

As mentioned in \S 4.4, $ ^{12} $CO 3--2 observations have also
been obtained at positions offset from the sources, in particular
for a 30$ '' $ offset (two JCMT beams). In all cases, the double-peaked
line profile disappears completely at the off position, confirming
that it arises from the circumstellar disk. In some sources,
however, a narrow profile at a velocity slightly offset from that
of the disk remains. This emission is due either to remnant envelope
material or the general molecular cloud from which the star formed.
Its strength is uncertain up to a factor of two since it was not possible
to find a good off-position in all cases. For the specific case of
LkCa 15, no emission was found at 30$ '' $ offset, but a weak $ ^{12} $CO
3--2 line with $ T_{\rm{MB}} $=0.22~K appeared at $ V_{\rm{LSR}}=-8 $
km s$ ^{-1} $ at the 1$ ' $ south position, where the H$ _{2} $
S(1) off-source spectrum was taken. This CO emission is more than
10 km s$ ^{-1} $ offset from the velocity of the star and is most
likely the result of a chance coincidence with a background cloud.

\subsection{$ ^{12}$CO 6--5 lines}

\begin{figure}
\figcaption{As Figure~6, but for CO 6--5 CSO spectra. \label{co_6_5}}
\plotone{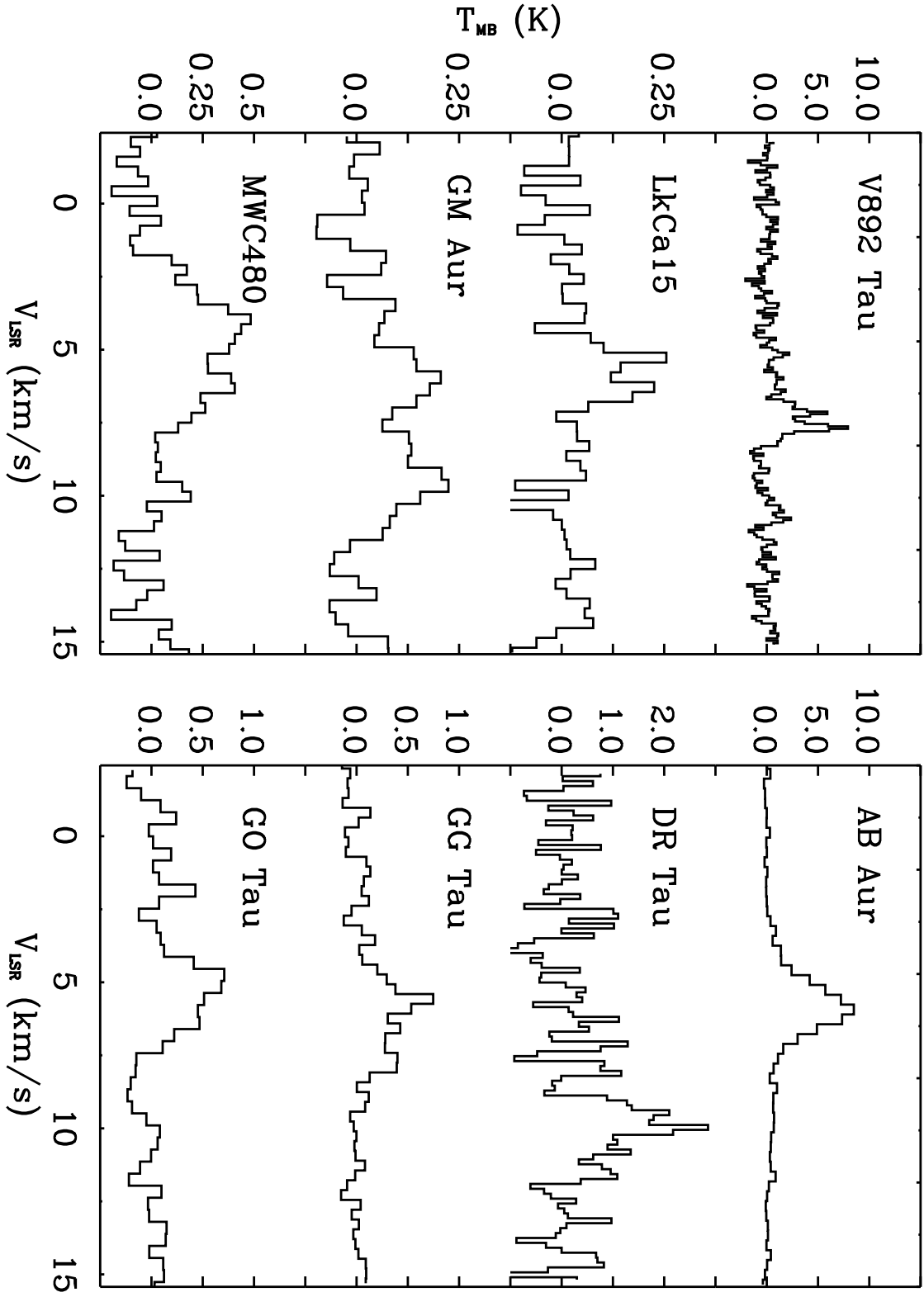}
\end{figure}

$ ^{12} $CO 6--5 emission is detected toward several sources using
the CSO (see Figure~\ref{co_6_5}). Weak, but clear double-peaked
profiles are seen from disks such as those around LkCa 15 and MWC
480 \citep[see also][]{GJ01}. The line is particularly strong toward
AB Aur, likely because of the extended envelope. Indeed, a small $ ^{12} $CO
3--2 and 6--5 map around the source shows strong lines even at one
beam offset. The integrated fluxes are reported in Table~\ref{co6_5}.

The 6--5 line probes preferentially gas at higher temperatures around
100 K, but its high optical depth decreases the excitation conditions
to lower temperatures. The ratio of the 6--5/3--2 line intensity is
a measure of this temperature \citep{GJ01}. A full analysis requires
a 2-D radiative transfer calculation for disk models with different
radial and vertical temperature profiles. However, a rough estimate
can be obtained from a simple 1-D escape probability formalism assuming
an iso-thermal and iso-density slab in which the abundances are chosen
such that the $ ^{12} $CO 6--5 and 3--2 lines are optically thick.
This slab would be representative of the intermediate and surface
layers of disks, from which most of the emission is thought to arise.

This simple analysis shows that the sources have a range of temperatures.
The upper limit on the CO 6--5 line for RY~Tau indicates a cool emitting
region of about 10~K. The sources LkCa~15, AB~Aur, GG~Tau and
GM~Aur all have relatively low temperatures between 20 and 50 K,
but these ranges can be extended considerably if typical calibration
errors of 20\% are taken into account, especially for GG~Tau and
AB~Aur. The sources GO~Tau, MWC~480, V892~Tau and DR~Tau have lower limits
to the temperatures of 30 K and in general suggest high temperatures
up to a few hundred Kelvin. Such high temperatures indicate that the upper 
layers of the disks are heated efficiently by the stellar light and are most
probably flared so that they capture the radiation far from the star.
Note however that the derived temperatures are extremely sensitive
to the errors in the line ratios. DR~Tau is surrounded by extended cloud
emission and the observed lines may well be emitted in different regions.

In summary, the combined detection of CO 6--5 and H$_2$ in 
several sources suggests that these sources may posses a warm upper layer,
consistent with a flared disk geometry. Two sources (RY~Tau and LkCa~15) 
could have lower 
temperatures on average which could either mean that the disk is flatter, or 
that dust-settling is taking place, reducing the heating of the gas in the 
upper layers of the disks \citep[e.g.][]{Chiang01}. Higher $S/N$ CO 6--5
 data and more accurate calibration are needed to use
the 6--5/3--2 ratio as an effective temperature probe (see \cite{GJ01} for 
a detailed discussion).

\subsection{Cold gas masses from CO}

Disk masses can be derived from the observed $ ^{13} $CO 3--2 data
assuming that most of the flux arises from the outer part of the disk
at a constant temperature. The simplification of an isothermal outer
disk is supported by detailed modeling along the lines of \citet{BS93}.
These models have a power-law decrease of the temperature with radius
to explain the behavior of the spectral energy distribution, but the
gradients in the outer disk are quite small. The reason could be that
the ambient interstellar radiation field incident on the outer disk
regulates the temperature structure with radius. Because of the large
beam dilution, our observations are not sensitive to the warm inner
gas, but only probe the outer cold gas. A common
outer gas temperature of 30~K is therefore assumed
for all our objects. This is slightly
higher than the temperature fixed by the local interstellar field,
which is around 10--15~K in quiet molecular cloud environments such
as found in Taurus and Ophiuchus. It is consistent with the observed
$ ^{13} $CO 3--2/1--0 line ratios \citep{GJ01}. In the optically
thin limit, the gas mass derived from $ ^{13} $CO 3--2 is given
by: \begin{equation}
M_{\rm{gas}}=3\times 10^{-6}\left( \frac{^{12}[\rm{C}]/[^{13}\rm{C}]}{60}\right) \left( \frac{\rm{H}_{2}/^{12}\rm{CO}}{10^{4}}\right) \frac{T_{\rm{ex}}+0.89}{e^{-16.02/T_{\rm{ex}}}}\frac{\tau }{1-e^{-\tau }}\left( \frac{d}{100\, \, \rm{pc}}\right) ^{2}\int T_{\rm{MB}}\, \, dV\, \, \rm{M}_{\odot }
\end{equation}
 The derivation of this formula is similar to that for CO 1--0 by
\citet{Sco86}. The mass varies by a factor of 2 for excitation temperatures
between 20 and 100~K, so that the exact value of the excitation temperature
is not crucial. Two main parameters must be assumed: the [$^{12}$C]/[$^{13}$C]
elemental isotope ratio and the H$_{2}$/$^{12}$CO conversion
factor. This factor is certainly not constant from source to source
and we adopt here a reference value of 10$^{4}$, typical for
dense molecular gas in which CO is not depleted.
As will be shown below, this factor is likely to be much larger
in disks due to the combined chemical effects of
freeze-out and photodissociation.

For sources for which no $^{13}$CO data are available (CQ Tau,
AA Tau, 49 Ceti, HD 135344), the $^{12}$CO data have been used
to determine the cold gas masses.

\subsection{Cold gas masses from millimeter continuum emission}

The cold gas masses can also be determined from the millimeter continuum
emission emitted by the cold dust, assuming a gas/dust ratio. All
sources in our sample have been previously observed in the millimeter
continuum with single-dish telescopes, usually at 1.3 or 1.1 millimeter
\citep[e.g.][]{BSCG90,OB95,MS97,MS00,Hen98,WSD89}. For some sources,
millimeter interferometer data exist at the same wavelengths, giving
similar flux levels \citep{MS97}, indicating that most of the single-dish
emission indeed comes from the disk rather than any remnant envelope.
To compute the cold dust mass, $ T_{\rm{dust}}=30 $~K is adopted,
similar to that found for CO. The value for the mass absorption coefficient
$ \kappa _{\lambda } $ (gas + dust) is taken to be 
$0.01(1.3\, {\rm{mm}}/\lambda ) $
cm$ ^{2} $ g$ ^{-1} $ from \citet{OH94} and 
assumes $ M_{\rm{gas}}/M_{\rm{dust}}=100 $.
The disk mass (gas + dust) is then given by: \begin{equation}
M_{\rm{disk}}=0.06\, \, {\rm{M}_{\odot }}\, \, \frac{F_{\lambda }}{1\, \, \rm{Jy}}\left( \frac{d}{100\, \, \rm{pc}}\right) ^{2}\frac{50\, \, \rm{K}}{\langle T\rangle }\frac{0.01\, \, \rm{cm}^{2}\, \, \rm{g}^{-1}}{\kappa _{1.3\, \, \rm{mm}}},
\end{equation}
 where $ F_{\nu } $ is the observed flux at 1.3~mm in Jy. The
observational data and resulting masses are summarized in
Table~\ref{mmcont}. The errors in the observed fluxes are taken to be 
$\sim$30\%.

\section{Analysis}

\subsection{Comparison of derived masses}

In the previous section, we applied three methods to estimate the
masses of disks around pre-main-sequence and debris-disk stars, summarized in 
Table~\ref{summary}. The derived masses differ considerably, well beyond the
error bars. We now discuss the strengths and weaknesses of each of these 
methods.

\begin{figure}
\figcaption{Estimated gas mass obtained from CO observations (upper panel) 
and H$_2$ observations (lower panel) plotted against the disk mass 
computed from 1.3mm dust assuming a gas/dust ratio of 100:1. 
In the upper panel, the dashed lines separate the regions of different 
CO depletion factors. \label{co_H2_dust}}
\plotone{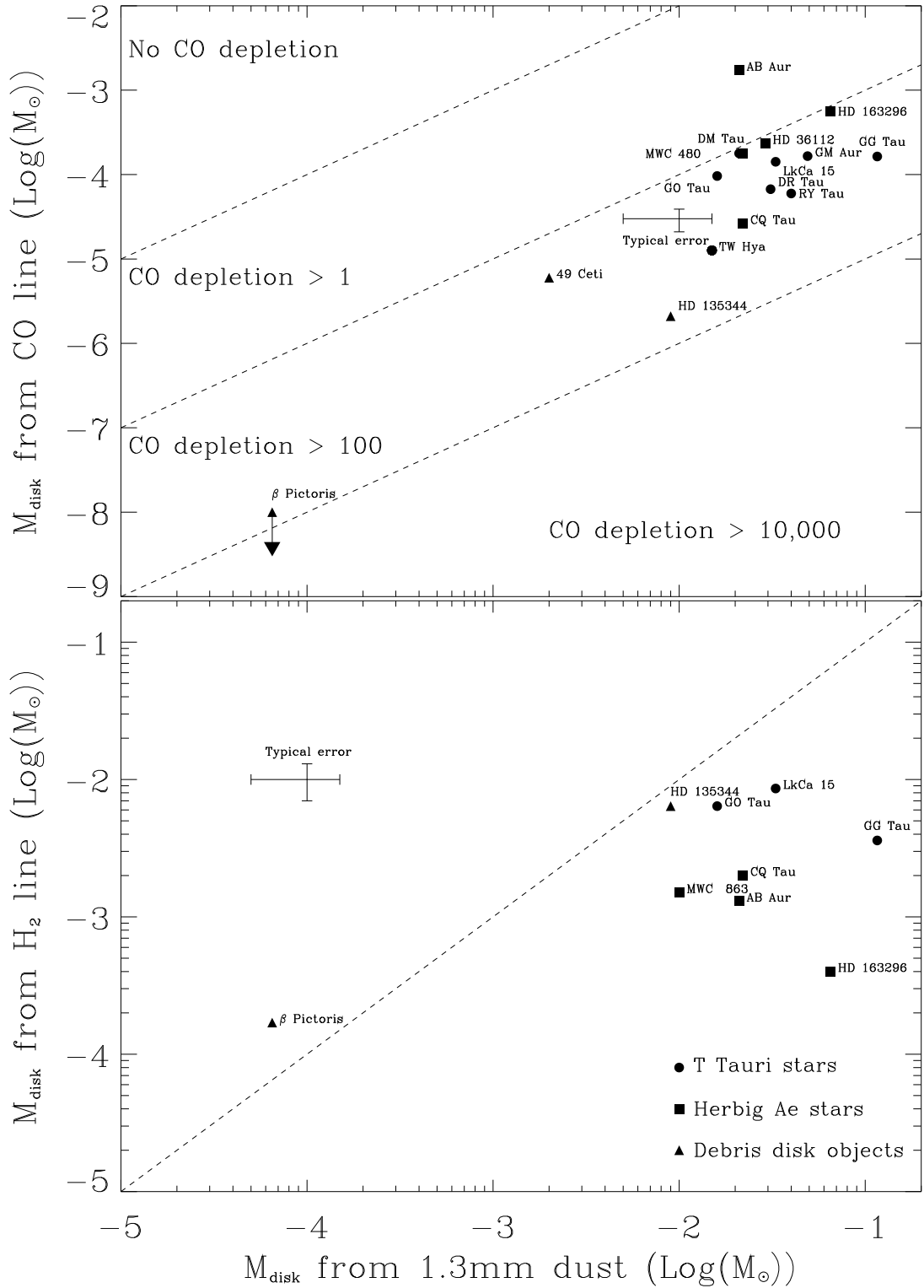}
\end{figure}

In the upper panel of Figure~\ref{co_H2_dust}, the masses obtained
from the $ ^{13} $CO 3--2 spectra are compared to those computed from
the 1.3 millimeter continuum emission assuming a mean disk temperature
of 30~K for the T~Tauri and Herbig~Ae stars. The dust around HD~135344
and $ \beta $~Pictoris has been taken to be warmer at 95 and 85~K
respectively \citep[see][]{CW95,Dent00}. The results for sources for
which only $ ^{12} $CO data are available are included, as well as
those for TW Hya studied by \citet{GJ01}. As found in previous studies
based on lower-$ J $ transitions \citep[e.g.][]{Dut96,MS97,MS00}, the
masses derived from CO are in general factors of 10--200 lower than
those found from the millimeter continuum. No distinction can be made
between T~Tauri and Herbig~Ae stars. The debris-disk objects as well as
TW~Hya seem to suffer very strong CO depletion, more than a factor of
10$ ^{3} $, in agreement with previous studies
\citep[e.g.][]{LA98,Dent95}.  Many explanations have been put forward,
including depletion of CO onto grains and dispersal of the disk
gas. As argued in \S 5.1, optical depths effects are unlikely to be
the main cause. \citet{GJ01} show that the
underabundance of CO is plausibly caused by a combination of
freeze-out in the coldest regions of the disk near the mid-plane, as
well as photodissociation of CO in the upper layers of the disk by
stellar and interstellar ultraviolet radiation
\citep{Yuri96,Inga00,WL00}. Substantial depletions due to freeze-out have
been found in dense, cold molecular cloud cores and in young stellar objects
environments \citep[e.g.][]{Kramer99,Shuping01}.

The millimeter continuum method is not exempt from difficulties either.  
In particular, it suffers from the
poor knowledge of the dust opacity constants
$\kappa_{\lambda}$. Theoretically, $\kappa_{\lambda} $ should be well
determined for particles that are much smaller than the wavelength of
observation, but its value depends strongly on the assumed particle
composition (silicates, amorphous carbon, ice mantle) and also on
particle size, shape and fluffyness. Within the range of possible
values, however, $\kappa_{\lambda}$ remains small enough to ensure
that the emission is optically thin beyond a few AU, so that the
determination of the total mass is quite straightforward.  Other
assumptions in this method include a constant $\kappa_{\lambda}$ for
the whole sample (i.e., no evolution of the opacity constant) and the
gas/dust ratio of 100:1 in the disks.

In the lower panel of Figure~\ref{co_H2_dust}, the warm gas masses
derived from the H$_{2}$ lines are plotted as functions of the
total gas masses obtained from the 1.3 millimeter continuum. For the
pre-main sequence stars, the warm gas masses are a fraction (1--10\%)
of the total gas masses, assuming a gas/dust ratio of 100. Some sources
such as LkCa 15, however, show a much larger fraction, of order 30\%.
\citet{Chiang01} modeled the Spectral Energy Distribution of LkCa 15 and
concluded that this source shows the strongest vertical dust settling. 
In the case of GO Tau, contamination by surrounding emission is possible. 
The warm gas masses have also been plotted versus the cold gas masses derived
from $^{13}$CO, but no correlation is found, as expected.

\subsection{Age determination and evolutionary trends}

In order to search for evolutionary trends in our results, the ages of
the stars need to be known. This is usually done by comparing the
positions of the stars on a Hertzsprung-Russell diagram with
theoretical evolutionary tracks. These tracks have many implicit
assumptions, however, and give different results depending on the
choice of the equations of state, the model used for convection, the
opacities, etc. \citep[see][for a review]{CB00}. On the observational
side, there are also uncertainties in the distance estimates of the sources,
the extinction and to what extent the intrinsic luminosity is affected by disk
accretion. The precise spectral type of few sources like MWC~480
remains controversial: A4 or A3ep+sh according to \cite{SDG00} and
\cite{MS97} respectively.
Moreover, all T~Tauri stars exhibit photometric variability, preventing
a precise determination of their characteristics. Some stars such
as RY~Tau and GG~Tau are binary systems and thus their stellar characteristics
must be corrected.

\begin{figure}
\figcaption{The evolutionary tracks of \citet{SFB00} for a metal
abundance Z=0.02. The left panel corresponds to intermediate mass stars
(2--3 M$_{\odot}$) and the right panel to low mass stars (1--2
M$_{\odot}$).  The location of our sources are overplotted.  The
different tracks correspond to the masses indicated next to each
track; the tickmarks along each track indicate the ages in Myrs.
\label{HR_siess}}
\plotone{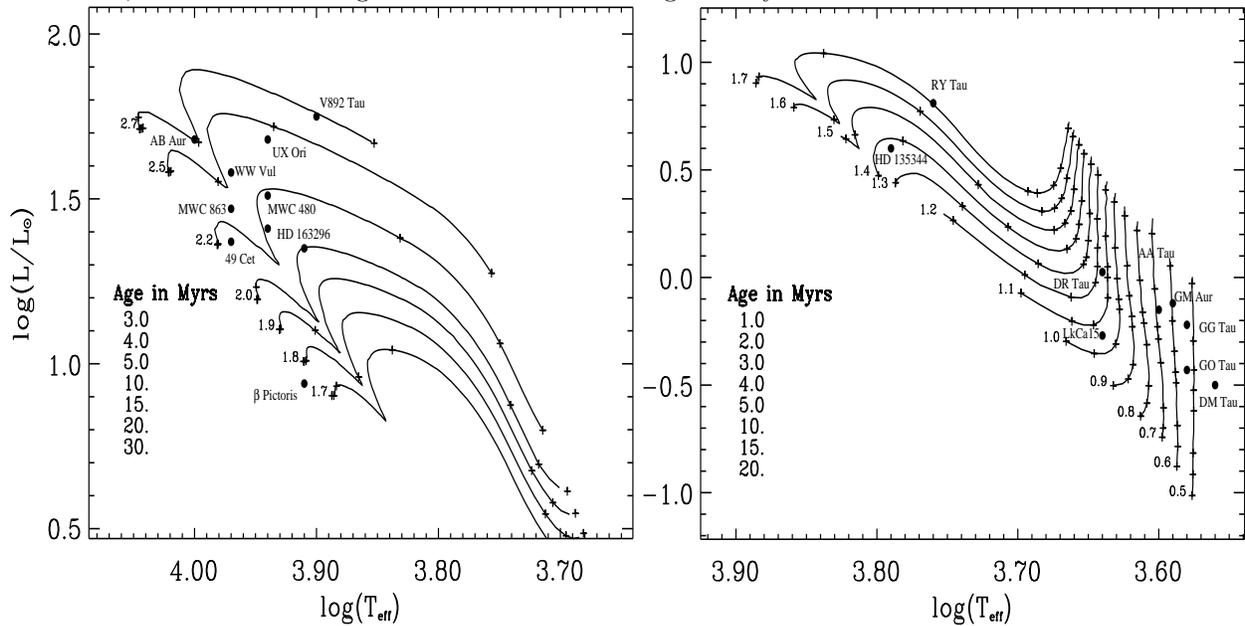}
\end{figure}

Although these factors result in significant absolute uncertainties,
the relative ages may be less affected. To obtain a consistent set of
relative ages, we have re-estimated the ages of the stars in our
sample using the recent pre-main-sequence evolutionary models of
\citet{SFB00}, which take the accretion history into account.  The
results are shown in Figure~\ref{HR_siess}. We take any binary systems
to be single stars, so that their ages should be considered rough
estimates.  The newly evaluated ages are consistent with previous
determinations and are listed in Table~\ref{ages}. If the tracks
of \citet{DM97} are used, a similar age ordering is obtained. The
discrepancies are largest for brown dwarfs and stars younger than
10$^{6}$ years.  Since our stars have higher mass ($>$0.5 M$_{\odot
}$) and ages greater than one million years, the differences between
the models are not significant. It is not the purpose of this paper to
discuss the validity of the different tracks.  The errors 
in our derived ages are of order
1--2 millions years, increasing for the older objects. The ages of the
intermediate mass stars are less well determined because their
effective temperature and luminosity do not vary significantly over a
large range of ages.  In particular, the age of $\beta$~Pictoris is
controversial.  Recently, \cite{BSS99} argue that $\beta$~Pictoris is
only (20$ \pm $10) $ \times 10^{6} $ years old with the error bar
reflecting the uncertainties in the isochrones used to derive the
age. The young age of $\beta$~Pictoris is consistent with the view
that it is part of a cluster of recent nearby star formation
\citep[e.g.][]{ZW00}. Whatever its actual age, $\beta$~Pictoris is the
oldest member in our sample.

\begin{figure}
\figcaption{Variation of the total disk mass with the age of the
central object deduced from the three methods.\label{mass_age}} 
\plotone{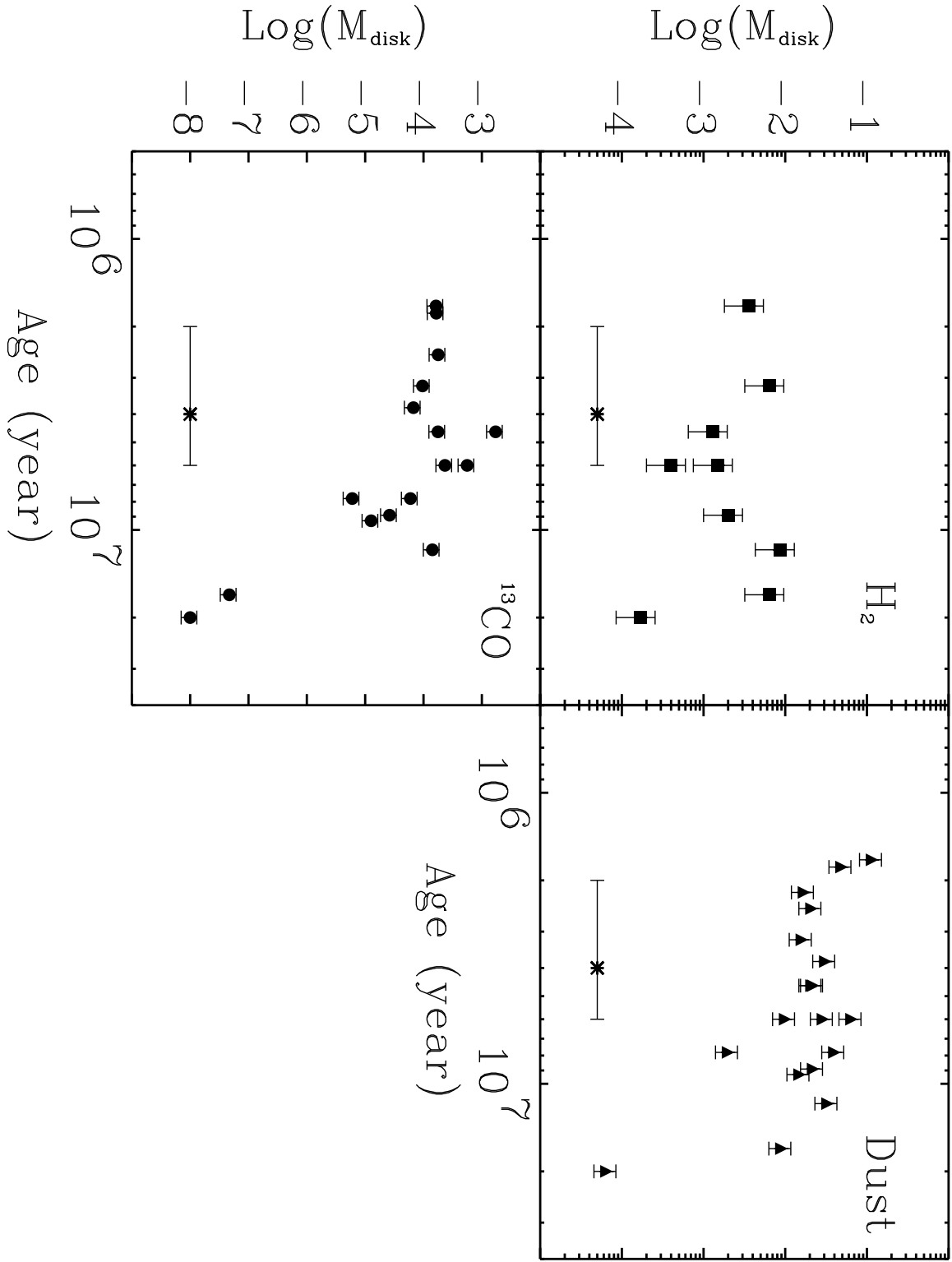}
\end{figure}

\begin{figure}
\figcaption{Evolution of the ratio of the total gas mass to the solid mass 
in circumstellar disks. The standard interstellar ratio is 100:1. 
\label{gas_to_dust}}
\plotone{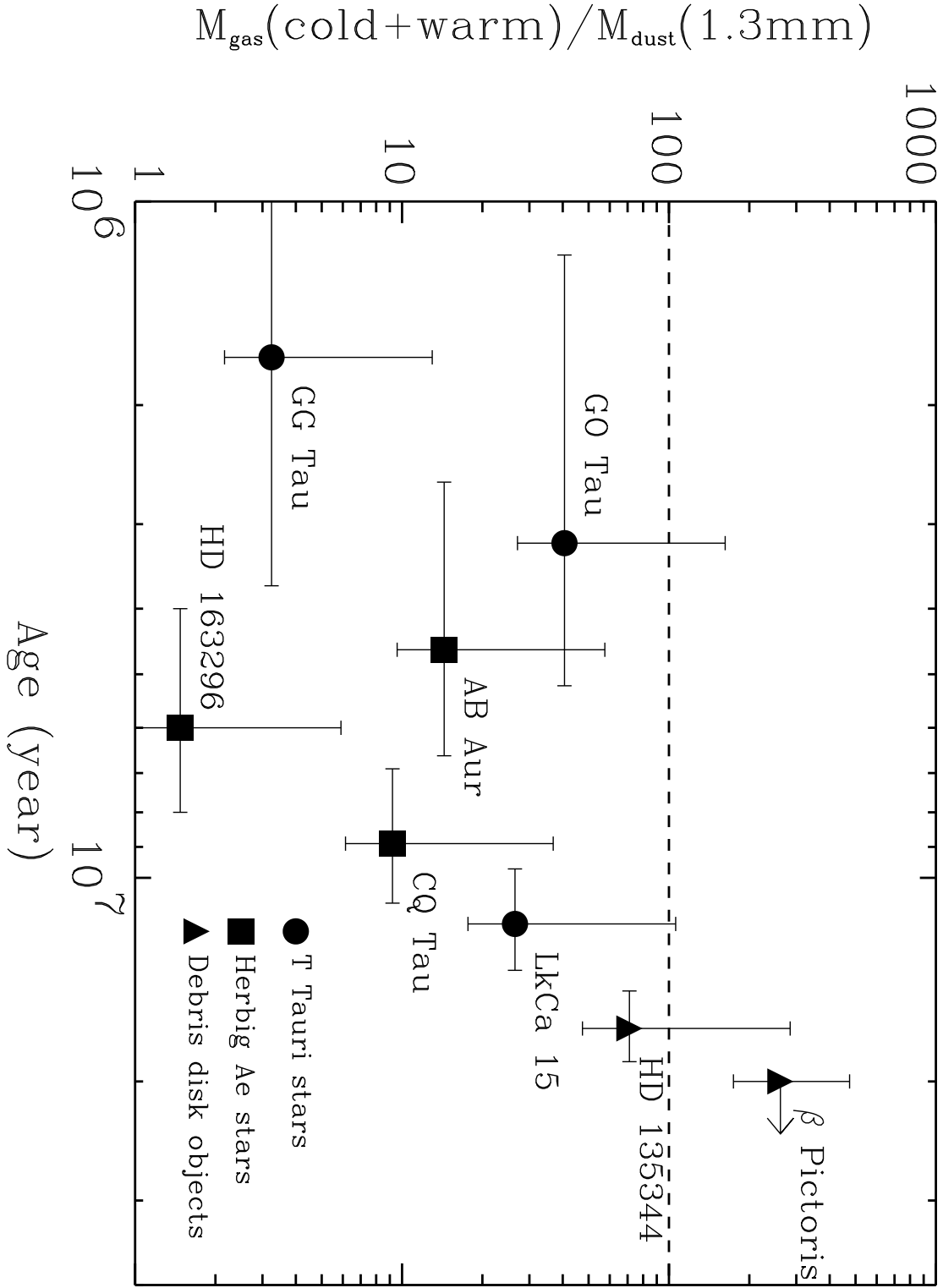}
\end{figure}

Figure~\ref{mass_age} shows the total disk masses deduced from the
three methods plotted against the ages of the stars. No strong
evolutionary trend appears but the behavior seems to be similar for
the three methods.  Figure~\ref{gas_to_dust} presents the total warm +
cold gas masses derived from the H$_{2}$ + CO data relative to the
total dust mass derived from the 1.3 millimeter continuum versus
age. As discussed in \S 6.4, only the debris-disk objects show a
gas/dust ratio close to 100:1, but this may be coincidental; for the younger
objects, a significant amount of cold H$_{2}$ is likely present, but
is not traced by CO.

Care has to be taken in the interpretation of these data, however.  As
mentioned before, the choice of objects in our sample is biased toward
the higher disk masses and some of the detections are marginal.
In fact, so-called weak-line T~Tauri stars are
surrounded by disks with lower masses \citep{Brandner00}.  This is
consistent with the non-detection of H$_{2}$ in these objects by
\citet{Stap99}. Our data are not sensitive to masses as small as
$10^{-4}$ M$_{\odot}$ for objects at a distance of 140 pc. Similarly,
$\beta$~Pictoris may be unusual since it is one of the dustiest
members of the debris-disk family.  Finally, it is difficult to compare
different absolute masses since the mass of the disk at a given time
of its evolution likely depends on the initial mass available.

\subsection{Heating mechanisms}

The derived amount of warm gas is significant and raises the question
of the source of heating. \citet{WF99a} discussed several
possibilities, including photon heating by stellar and interstellar
radiation and shock-heating caused by the interaction between a
stellar wind and the surface of disks. Here we investigate whether the
observed trends provide further clues to the dominant
mechanisms. Quantitative discussions and detailed modeling are left
for future work.

\begin{figure}
\figcaption{Panels (a) and (c) show the warm gas mass derived
from H$_2$ and the corresponding H$_2$ 
excitation temperature as a function of the effective temperature of the 
central star. Panels (b) and (d)
plot the variation of the warm gas mass and excitation 
temperature 
against the continuum flux at 28 $\mu$m normalized at 100~pc.
\label{corr4}}
\plotone{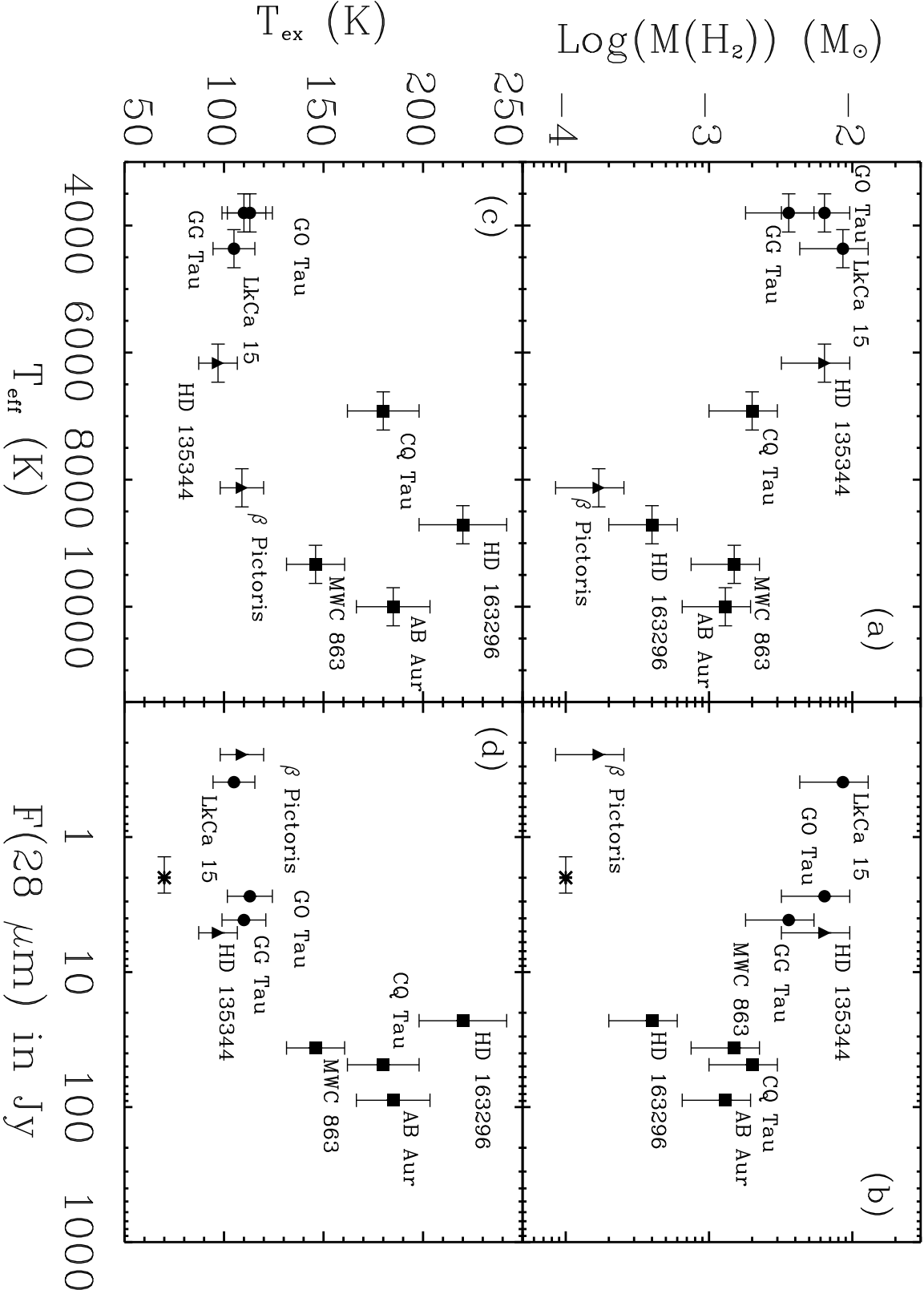}
\end{figure}

Since the disks in our sample have negligible accretion onto the star
(typically $<$ 10$^{-8}$ M$_{\odot }$ yr$^{-1}$), the irradiation of
the central object should control, at least partially, the temperature
profile of the disks. To study this scenario, we plot in
Figure~\ref{corr4} the excitation temperatures derived from the
H$_{2}$ S(0) and S(1) lines as functions of the effective temperature
of the star. Obviously, no significant correlation is found in this
figure. We can, however, distinguish three groups: T~Tauri, Herbig~Ae
and debris-disk stars. The T~Tauri stars have gas at $\sim$100~K,
whereas their higher mass counterparts are surrounded by gas at 150~K
or more. The higher $T_{\rm{ex}}$ observed in disks around Herbig Ae
stars suggests that the harder stellar ultraviolet radiation can be
transformed more efficiently into heat for these objects. However, if
the number of photons with wavelengths
$<$1100 \AA \ capable of ionizing atomic carbon
is also increased, this results in a larger C$^{+}$ abundance,
increasing the cooling as well. Detailed modeling of the surface
heating as a function of radiation field is needed. Note that
classical models of photon-dominated-regions (PDRs) do not show any
increase of the H$_{2}$ excitation temperature versus strength of the
incident radiation field for the normal interstellar field typical of
a B0 star, even though the surface temperatures
increase \citep[e.g.][]{Mario00a}. However, the variation of
$T_{\rm{ex}}$ with effective temperature of the star has not yet been
modeled.

A link seems to exist between the excitation temperature and the
continuum flux at 28 $\mu$m normalized at 100~AU
(Figure~\ref{corr4}$d$).  Above a certain threshold, the excitation
temperature increases as the continuum flux becomes higher. As a
consequence, the warm gas mass drops with continuum flux or
$T_{\rm{eff}}$ because of the steep dependence of the mass on
temperature (Eq. (3) and Figure~\ref{masstemp}) (see
Figure~\ref{corr4}$a$ and \ref{corr4}$b$). Moreover, the fraction of
warm gas to the total gas mass around typical A stars like HD~163296
or AB~Aur is small compared to that around T~Tauri stars.

The role of ultraviolet radiation in heating the surfaces of flared
disks is taken into account in recent models by \citet{CG97,CG99},
\citet {DA98} and \citet{Bell97}. As shown by \citet{WF99a}, these
models fall short of explaining the observed masses of warm H$ _{2} $
gas by factors of at least a few. It is not yet clear whether this
discrepancy is significant, since the same models also fail for normal
molecular clouds unless the grain formation rate of H$ _{2} $ is
significantly enhanced \citep[e.g.][]{Harbart00,Li01}. The presence of
a thin envelope can enhance the scattered stellar radiation and thus
also the warming \citep{Natta93}. At the edges of PDRs, the main
heating agent is the photoelectric effect on grains, with small grains
and PAHs being particularly effective
\citep{HT97,BT94}. \citet{Spaans94} investigated the influence of the
effective temperature of the central illuminating star on the gas
heating efficiency by very small grains (grain radius between 4 and
180 \AA) and PAHs. They showed that the efficiency drops only by a
factor of 4 from a star at 10000~K to one at 4000~K. Adding the fact
that most T~Tauri stars exhibit ultraviolet excess and a strong Lyman
alpha emission line, the effective heating by low mass stars compared
to intermediate mass stars should be similar.  The detections of PAHs
around AB~Aur \citep{Mario00b} and HD~135344 \citep{CWD98} suggest
that these large molecules can play a role in the heating of the
disks, but quantitative models have not yet been performed. The gas
can attain higher temperatures than the dust in these layers,
consistent with our observations, and its emission can emerge from the
surfaces even if the mid-plane is optically thick in the mid-infrared
continuum. The efficiency of photoelectric heating decreases significantly,
however, if the size of the grains is increased, so the dust size distribution
also plays an important factor in this analysis \citep{KvZ01}.

Alternatively, the line emission can escape through `holes' or `gaps'
in the disk created by low-mass companion(s), e.g. planets or brown dwarfs
\citep{LubowArt00}.
Such gaps could also result in a larger surface area
intercepted by the radiation. In any case, the detection of
ultraviolet emission from fluorescent H$_{2}$ toward other
pre-main-sequence stars indicates that ultraviolet radiation plays
some role in these systems \citep{VJL00}. Note that this fluorescent
H$_2$ seen in the ultraviolet must arise from much hotter gas, of
order 2000 K, probably located in an inner boundary layer close to the
star.

Possible heating of H$_{2}$ by shocks created by the interaction of a
stellar wind with the surface of disks was discussed by \citet{WF99a}.
A significant constraint is however provided by the non-detection of
the H$_{2}$ S(3) lines in our sources, since shocks tend to populate
the high-$J$ H$_{2}$ levels as well. Shock models by \citet{BHT92}
give much higher H$_{2}$ excitation temperatures than observed, making
them less plausible.


\subsection{Gas content of debris disks}

The most interesting cases are formed by the debris disk objects HD
135344, $\beta$~Pictoris and 49 Ceti. The disks around these objects
are considered gas poor based on CO observations assuming a H$_{2}$/CO
conversion factor of 10$^{4}$ \citep{WF01}. For HD 135344 and $\beta
$~Pictoris, two lines are possibly detected (in more than 50\% of the
spectra obtained by our data reduction procedure), giving a measure of
the temperature and the mass with 10\% and 55\% uncertainty,
respectively, if only the standard 30\% calibration errors are
considered.  For 49~Ceti, only the S(0) line is seen, leading to an
H$_{2}$ mass of (3.5$\pm$1.9)$ \times $10$^{-3}$ M$_{\odot}$ if
$T_{\rm{ex}}=100$~K is assumed. The derived mass of gas around
$\beta$~Pictoris is $\sim 10^{-4}$ M$_{\odot}$ or $ \sim$ 0.17$ \pm
$0.09 M$_{J}$, and it is $\sim$ 6.4 $10^{-3}$ M$_{\odot}$ around
HD~135344. The amount of gas in the $\beta$ Pictoris disk is
significantly smaller than that for other disks. The disks are not
resolved within the ISO-SWS beam for HD 135344 and 49 Ceti and barely
resolved for $\beta$~Pictoris. Therefore, the location of the emitting
gas is unknown.

The detection of H$_2$ gas in the $\beta$~Pictoris disk may seem
surprising since this disk has a very low CO/dust mass ratio
\citep[e.g.][]{LA98, Zuc95, Roberge00}. The presence of some neutral
gas was, however, invoked by \citet{Lagrange98} in order to slow down
ions like \ion{Ca}{2} or \ion{Fe}{2} leaving the disk since these ions
suffer strong radiation pressure from the star.
They considered only \ion{H}{1} as a major 
species of the stable ring.
The detection of \ion{Fe}{2} lines together with the measured 
\ion{Fe}{1}$/$\ion{Fe}{2} ratio implies densities of 10$^{3}$--10$^{6}$
cm$ ^{-3} $ \citep{Lagrange95}. This is 
consistent with the density at 40 AU in the disk model
of \citet{Inga00} which has M$_{\rm{disk}}$ $\sim$ 10$^{-4}$ M$ _{\odot } $,
similar to that found here.  In their model of debris disks, \citet{Inga00}
consider the balance of the formation and destruction
of CO and H$_2$. Their main conclusion is that the CO molecule can exist
only in the dense part of disks protected from photodissociation 
whereas H$_{2}$ is widely spread. Moreover, CO freezes out onto dust 
in the coldest parts near the midplane, making it a poor tracer of the gas.


If the H$ _{2} $ detections are valid, nearly all of the gas is at
high temperatures ($ \sim 80-100 $ K) in the debris disks.  Since
the disk of $ \beta $~Pictoris has an optical depth less than unity at
optical wavelengths, the ultraviolet photons can warm the gas in the
entire disk through the photoelectric effect and other processes
\citep[]{KvZ01}. These gas temperatures are only slightly higher than those
derived for the dust components: the spectral energy distributions of
$\beta$~Pictoris and HD 135344 are well fitted by a single dust
temperature of 80--90~K \citep[]{WH00,CW95}, indicating that these
disks are globally warmer than those around T~Tauri or Herbig Ae stars. 

The estimated total dust mass around $\beta$~Pictoris ranges from
0.3$\times$10$^{-6}$ M$_{\odot}$ \citep[]{WH00} to
10$^{-6}$ M$_{\odot}$ \citep[]{LG98}. The gas-to-dust mass ratio lies
therefore between 45 and 380, and is much higher than the value of
0.02 derived from CO ultraviolet observations.  Note that if these CO
molecules are the evaporation products of infalling comets onto
$\beta$~Pictoris \citep[]{Lecavalier96}, it is not possible to derive
the primordial H$_2$ content of the disk from CO.

A pertinent consequence of the presence of gas in debris disks is that
it affects the dynamics of the dust in those disks
\citep[e.g.][]{W77}.  For sufficiently large gas masses, dust
generation by collisions of planetesimals will not be possible.  For
$\beta$ Pictoris, however, the gas mass of $\sim$ 0.1 M$_{J}$ is small
enough that it does not prevent a collisional cascade. \cite{TA01}
modeled the evolution of dust grains in disks with gas masses up to a few
tens of Earth masses, comparable to that
found for $\beta$~Pictoris. They show that although grains migrate radially
due to radiative pressure and gravity, equilibrium orbits exist for a
specific range of grain sizes. Most interestingly, their models can
reproduce ring-like disk morphologies with an inner disk clear of small
grains. 
The similarity of our derived gas/dust ratio for $\beta$ Pictoris
with the interstellar value of $\sim$100:1 is therefore likely coincidental:
some of the dust and gas may have accumulated
into (gaseous) planets and planetesimals, been expelled from the disk
due to radiation pressure, or fallen onto the star by
Poynting-Robertson drag.

\section{Conclusions}

We have conducted the first survey of H$_{2}$ rotational line emission
from disks around a sample of T~Tauri and Herbig~Ae stars and 
from young stars with debris disk using the ISO-SWS. 
The observed spectra reveal the presence of an
unexpectedly large amount (0.1--10 $\times$ 10$ ^{-3} $ M$_{\odot}$) of
molecular gas at $\sim$100~K. No correlations between the warm gas
masses with disk masses derived from $ ^{13} $CO and 1.3 millimeter
emission were found. Whereas the bulk of the gas around T~Tauri and
Herbig Ae stars is cold, the warm gas may constitute the major gaseous
component of debris-disk objects like HD~135344 and
$\beta$~Pictoris. There is no apparent difference between the low and
the intermediate mass pre-main-sequence stars. The possible heating mechanisms
responsible for the warm gas are discussed. No process can adequately
explain the large amount of warm gas, but the ubiquitous presence of
warm H$ _{2} $, the higher gas than dust temperatures, and the
detection of PAHs in few of the objects suggest that a common
mechanism like photoelectric heating by ultraviolet radiation could be
the main heating agent. Further modeling is needed.

Complementary observations of $^{12}$CO 3--2, 6--5 and $^{13}$CO 3--2 have been
performed. The line profiles are resolved and exhibit double-peaked
features consistent with gas emitted from a disk in Keplerian rotation
around a central object. Ratios of integrated fluxes of the two
isotopomers $^{12}$CO and $^{13}$CO show that the $^{13}$CO 3--2 line
is not highly optically thick and potentially a tracer of the cold
component of disks. The presence of warm gas is supported by the
detection of $^{12}$CO 6--5 toward a few sources where H$_2$ has also been 
found. The gaseous masses inferred from the $^{13}$CO intensities
are much smaller than those found from the dust continuum emission.
CO is likely strongly affected by photodissociation via the
stellar and interstellar ultraviolet radiation in the surface
layers and freeze-out onto grain surfaces in the midplane.

The H$_{2}$, CO and millimeter continuum data together with rough age
estimates of our stars allow evolutionary trends to be investigated.
No strong evolution in the masses derived from CO, H$_{2}$ or dust is
found. There is a large diversity among the stars studied in the
(1--10)$\times 10^{6}$ years range. The limited number of objects, the 
limited quality of the ISO data and uncertainties in the derived masses
prevent definitive conclusions on the gas survival time scale.

The analysis of the H$_{2}$ data presented here suffers greatly from
limited spatial and spectral resolution as well as sensitivity.
Ground-based spectrometers soon to be operational on 8--10 m class
telescopes will be able to study the S(1) line at vastly higher
spectral and spatial resolution, but will not have access to the
ground state para-H$_{2}$ transition at 28 $\mu$m. Moreover, the
surface brightness sensitivities of these warm large telescopes is
only marginally improved compared with small cryogenic space
observatories such as ISO. More complete studies with future air- and
space-borne mid-infrared spectrometers on SIRTF, SOFIA, and eventually
NGST will greatly improve on our ability to examine the H$_{2}$
emission lines from young stars, and properly address the many
interesting questions associated with the structure of circumstellar
disks and the formation of giant gaseous planets raised in this paper.

\acknowledgements{}

This work was supported by the Netherlands Organization for Scientific
Research (NWO) grant 614.41.003 and a Spinoza grant , and by grants to 
GAB from NASA (NAG5--8822 and NAG5--9434). A.N. is supported in part by ASI
ARS-98-116 grant. Discussions with Eugene Chiang, Peter Goldreich, Michiel
Hogerheijde, Frank Shu and Doug Johnstone are appreciated.  
The authors thank the staff of the CSO and JCMT, in particular Fred Baas 
and Remo Tilanus, for their support and the Dutch ISO Data Analysis Center
(DIDAC) at SRON-Groningen, especially Edwin Valentijn and Fred Lahuis, 
for their help during the data reduction of the ISO-SWS spectra.


\clearpage
\begin{deluxetable}{llllllll}
\tablewidth{0pt}
\tablecaption{Stellar characteristics\label{sources}}
\tablehead{\colhead{Name}&\colhead{SpT}&\colhead{$\alpha$(J2000)}&\colhead{$\delta$(J2000)}&\colhead{$\log\left(T_{\rm{eff}}\right)$}&\colhead{$\log$(L$_{*}$/L$_{\odot})$}&\colhead{$d$\tablenotemark{a} (pc)}&\colhead{Ref.}}
\startdata   
\multicolumn{8}{c}{T~Tauri stars}\\
\\
AA~Tau                   & K7   & 04 34 55.5  & $+$24 28 54   &  3.60     & $-$0.15    & 140 & 1 \\ 
DM~Tau                   & M0.5 & 04 33 48.7  & $+$18 10 12   &  3.56     & $-$0.5     & 140 & 1 \\
DR~Tau                   & K7   & 04 47 06.3  & $+$16 58 41   &  3.64     & $+$0.025          & 140 & 2\\
GG~Tau\tablenotemark{b}  & K7   & 04 32 30.3  & $+$17 31 41.0 &  3.58    & $-$0.22 $\pm$0.23 & 140 & 3    \\
GO~Tau                   & M0   & 04 43 03.1  & $+$25 20 19   &  3.58     & $-$0.43           & 140 & 1 \\
RY~Tau\tablenotemark{c}  & K1   & 04 21 57.41 & $+$28 26 35.6 &  3.76    & $+$0.81           & 133   & 2, 4   \\
GM~Aur                   & K7   & 04 55 10.2  & $+$30 21 58   &  3.59     & $-$0.12          & 140 & 2  \\      
LkCa~15                  & K7   & 04 39 17.8  & $+$22 21 03   &  3.64     & $-$0.27          & 140  & 2    \\

\cutinhead{Herbig~Ae stars}
UX~Ori    & A3IIIe  & 05 04 29.9  & $-$03 47 14.3 &  3.94 & $+$1.51 $^{+0.15}_{-0.13}$ &  430 & 7\\
HD~163296 & A3Ve    & 17 56 21.26 & $-$21 57 19.5 &  3.94 & $+$1.41 $\pm 0.69$ & 122 $^{+17}_{-13}$ & 8\\
CQ~Tau    & F5IVe   & 05 35 58.47 & $+$24 44 54.1 &  3.84 & $-$0.21 $^{+0.19}_{-0.16}$ & 100 $^{+25}_{-17}$ & 8\\
MWC~480   & A3ep+sh & 04 58 46.27 & $+$29 50 37.0 &  3.94 & $+$1.51 $^{+0.15}_{-0.13}$  & 131 $^{+24}_{-18}$ & 8\\
MWC~863   & A1Ve    & 16 40 17.92 & $-$23 53 45.2 &  3.97 & $+$1.47 $^{+0.25}_{-0.19}$   & 150 $^{+40}_{-30}$ & 8\\
HD~36112  & A5IVe   & 05 30 27.53 & $+$25 19 57.1 &  3.91 &  $+$1.35 $^{+0.24}_{-0.18}$   & 200 $^{+60}_{-40}$ & 7\\
AB~Aur    & A0Ve+sh & 04 55 45.79 & $+$30 33 05.5 &  4.00 & $+$1.68 $^{+0.13}_{-0.11}$  &144 $^{+23}_{-17}$ & 8\\
WW~Vul    & A0      & 19 25 58.75 & $+$21 12 31.3 &  3.97 & $+$0.73 & 550  & 9\\
V892~Tau  & A0      & 04 18 40.61 & $+$28 19 16.7 & 3.90  & $+$1.75   & 140   & 10\\
\cutinhead{Debris-disk stars}
49~Ceti           &  A1V & 01 34 37.78 & $-$15 40 34.9 &  3.97 &  $+$1.37     & 61 & 11\\
HD~135344        &  F8V & 15 15 48.44 & $-$37 09 16.0 &  3.79 &  $+$0.60        & 80 & 12\\
$\beta$~Pictoris &  A5V & 05 47 17.09 & $-$51 03 59.5 &  3.91 & $+$0.94      & 19.28 $\pm 0.19$& 13\\
\enddata
\tablenotetext{a}{In cases where no accurate (Hipparcos) distance is
available, a mean distance of 140~pc is adopted \citep{Kenyon94}.}
\tablenotetext{b}{Characteristics of the most massive star of the binary system.}
\tablenotetext{c}{Possible binary system \citep{Bertout99}.}
\tablerefs{(1) \citealt{Hart98};
(2) \citealt{SFB99};
(3) \citealt{Ghe97};
(4) \citealt{Wich98};
(5) \citealt{KH95};
(6) \citealt{Webb99};
(7) \citealt{Natta99};
(8) \citealt{Mario97};
(9) \citealt{FRGT93}; 
(10) \citealt{BCILNS92};
(11) \citealt{CWD98};
(12) \citealt{CW95}
(13) \citealt{Crifo97};
(14) \citealt{MS97}}
\end{deluxetable}

\clearpage
\begin{deluxetable}{lccccc}
\tablewidth{0pt}
\tablecaption{H$_{2}$ molecular line data \label{h2data}}
\tablehead{
\colhead{Transition}&\colhead{Wavelength\tablenotemark{a}}&\colhead{$E_{\rm{upper}}$\tablenotemark{a}}&\colhead{$A$--coefficient\tablenotemark{b}}&\colhead{$n_{\rm{crit}}$\tablenotemark{c} \  at 100~K}\\
   &\colhead{($\mu$m)}    &\colhead{(K)} & \colhead{(s$^{-1}$)} &\colhead{(cm$^{-3}$)}}
\startdata
H$_{2}$ S(0) 2$\to$0 & 28.218 & 509.88  & 2.94 10$^{-11}$&  54\\
H$_{2}$ S(1) 3$\to$1 & 17.035 & 1015.12 & 4.76 10$^{-10}$&  1.1 10$^{3}$ \\
H$_{2}$ S(2) 4$\to$2 & 12.278 & 1814.43 & 2.76 10$^{-9}$ &  2.0 10$^{4}$ \\
H$_{2}$ S(3) 5$\to$3 &  9.662 & 2503.82 & 9.84 10$^{-9}$ &  1.9 10$^{5}$\\
\enddata
\tablerefs{(a) \citealt{Jennings87}; (b) \citealt{Wol98}; (c) using H$_{2}$--H$_{2}$ 
collisional transition rate coefficients from \citet{Flower98}.}
\end{deluxetable}

\clearpage
\begin{deluxetable}{llllll}
\tablecaption{H$_{2}$ integrated fluxes with inferred temperature and mass \label{h2iso}}
\tablewidth{0pt}
\tablecolumns{6}
\tablehead{
\colhead{Name}& \colhead{H$_{2}$ S(0)}&\colhead{H$_{2}$ S(1)}& \colhead{$T_{\rm{ex}}$}&\colhead{H$_{2}$ mass} &\colhead{Level of}\\\colhead{}    & \multicolumn{2}{c}{(10$^{-14}$ erg s$^{-1}$ cm$^{-2}$)}&  \colhead{(K)} & \colhead{(10$^{-3}$ M$_{\odot}$)}&\colhead{confidence}}
\startdata
AA~Tau     &   $<$1.5       & \phm{$<$}8.1$\pm$0.25 & 100--200\tablenotemark{a} & 20.6--0.2 & medium\\
DR~Tau     &   $<$1.5       & $<$0.8 & \nodata                   & \nodata   & \nodata \\
GG~Tau     &   \phm{$<$}2.5$\pm$0.8 & \phm{$<$}2.8$\pm$0.8 & 110$\pm$11                       & 3.6$\pm$1.8   & high \\
GO~Tau     &   \phm{$<$}5.6$\pm$1.7 & \phm{$<$}7.1$\pm$2.1 & 113$\pm$11                       & 6.4$\pm$3.2   & medium\\
RY~Tau     &   $<$1.5       &  $<$0.8      & \nodata                   & \nodata   & \nodata \\
GM~Aur     &   $<$1.5       &  $<$0.8      &  \nodata                  & \nodata   & \nodata \\ 
LkCa~15    &   \phm{$<$}5.7$\pm$2.2 & \phm{$<$}5.3$\pm$1.6 & 105$\pm$10                       & 8.6$\pm$4.3   & medium\\
\\
\tableline
\\
UX~Ori     &  \phm{$<$}6.8$\pm$2.0  & $<$0.8             & 100--200\tablenotemark{a} & 117--9      & high \\
HD~163296  &  \phm{$<$}1.9$\pm$0.6  & \phm{$<$}22$\pm$6  & 220$\pm$22                & 0.4$\pm$0.2 & high \\
CQ~Tau     &  \phm{$<$}5.9$\pm$1.8  & \phm{$<$}40$\pm$12 & 180$\pm$18                & 2.0$\pm$1.0 & high \\
MWC~480    &  $<$1.5        & \phm{$<$}10$\pm$3  & 100--200\tablenotemark{a} & 78.8--0.7 & \nodata \\
MWC~863    &  \phm{$<$}6.9$\pm$2.1  & \phm{$<$}24$\pm$7  & 146$\pm$14                       & 1.5$\pm$0.8     & high \\
HD~36112   &  $<$1.5        & \phm{$<$}3.6$\pm$1.1 & 100--200\tablenotemark{a} & 18.7--0.2 & medium \\
AB~Aur     &  \phm{$<$}4.1$\pm$1.2  & \phm{$<$}30$\pm$9  & 185$\pm$18                       & 1.3$\pm$0.7     & high \\
WW~Vul     &  $<$1.5        &  $<$0.8      & \nodata                   & \nodata   & \nodata \\
\\
\tableline
\\
49~Ceti           &   \phm{$<$}6.6$\pm$2.0 &  $<$0.8      & 100--200\tablenotemark{a} & 2.3--0.3  & medium \\
HD~135344        &   \phm{$<$}9.0$\pm$2.7 & \phm{$<$}5.5$\pm$1.7 & 97$\pm$10                        & 6.4$\pm$3.2    & medium \\ 
$\beta$~Pictoris &   \phm{$<$}7.0$\pm$2.1 & \phm{$<$}7.7$\pm$2.3 & 109$\pm$11                       & 0.17$\pm$0.08  & medium \\
\enddata

\tablenotetext{a}{Assumed temperature range}
\tablecomments{All upper limits are 3$\sigma$.
No correction for extinction has been taken into account
in the calculation of the temperatures and masses. The errors on the
fluxes of $\sim$30\% translate into uncertainties of $\sim$10 \%
on the temperatures and $\sim$55\% on the mass. The level of confidence of
the detection is considered ``high" when the line is detected in more than 
75$\%$ of the 1000 reduced spectra. The level becomes
``medium" when the detection is present in about 50--75\% of the
spectra. In cases of non-detection, the line is seen in less than 50\%
of the reductions.}

\end{deluxetable}

\clearpage
\begin{deluxetable}{lrrrrrlc}
\tablecaption{Observed continuum fluxes near H$_{2}$ lines \label{mircont}}
\tablewidth{0pt}
\tablehead{
\colhead{Name}&\colhead{3.4 $\mu$m}&\colhead{6.9 $\mu$m}&\colhead{9.6 $\mu$m}
&\colhead{17 $\mu$m}&\colhead{28.2 $\mu$m}&\colhead{$T_{\rm{thin}}$}
&\colhead{Remarks}\\
\colhead{}&\colhead{(Jy)}&\colhead{(Jy)}&\colhead{(Jy)}&\colhead{(Jy)}
&\colhead{(Jy)}&\colhead{(K)}&\colhead{}}
 \startdata
AA~Tau           &    $<$0.1 &     $<$0.1 &        0.1 &        1.1 &        1.2 & \ 93$\pm$9 &\tablenotemark{a}\\
DR~Tau           &       2.0 &        1.7 &        2.4 &        4.3 &        5.4 & \ 90$\pm$9 &\\
GG~Tau           &       0.5 &        0.4 &        0.8 &        1.1 &        2.1 & \ 81$\pm$8 &\tablenotemark{a} \\
GO~Tau           &       0.2 &        0.2 &        0.3 &     $<$0.1 &        1.4 & \ 88$\pm$9 &\tablenotemark{a}\\
RY~Tau           &       5.2 &        5.1 &       15.9 &       17.0 &       16.9 & \ 92$\pm$9 &\\
GM~Aur           &    $<$0.1 &     $<$0.1 &        0.5 &        0.1 &        1.2 & \ 56$\pm$6 &\tablenotemark{a}\\
LkCa~15          &       0.3 &        0.4 &        0.5 &        0.4 &        0.2 & 120$\pm$12& \tablenotemark{a}\\
\\
\tableline
\\
UX~Ori           &       1.0 &        0.7 &        3.5 &        1.4 &        4.6 & \ 71$\pm$7  &\\
HD~163296        &       8.5 &        7.0 &       18.0 &       16.9 &       15.4 & \ 78$\pm$8  &\\
CQ~Tau           &       2.6 &        2.1 &        7.1 &       13.1 &       21.6 & 155$\pm$15 & \\
MWC~480          &       3.8 &        3.7 &        8.7 &        4.7 &        7.2 & \ 85$\pm$9   &\\
MWC~863          &       6.1 &        5.5 &       22.3 &       16.7 &       16.2 & \ 98$\pm$10   &\\
HD~36112         &       3.6 &        2.5 &        6.3 &        4.8 &        6.4 & \ 81$\pm$8   &\\
AB~Aur           &      13.2 &        9.6 &       29.9 &       24.4 &       45.4 & \ 96$\pm$10   &\\
WW~Vul           &       0.9 &        0.7 &        2.3 &        1.8 &        2.1 & \ 84$\pm$8   &\\
\\
\tableline
\\
49~Ceti           &       1.8 &        0.6 &        0.3 &        0.8 &        0.2 & 161$\pm$16 &\tablenotemark{a}\\
HD~135344        &       3.2 &        2.1 &        1.3 &        2.8 &        8.0 & \ 74$\pm$7  &\\
$\beta$~Pictoris  &      12.4 &        3.4 &        2.7 &        3.0 &        6.6 & \ 96 $\pm$10 &\\
\enddata
\tablecomments{Photometric errors are $\sim$ 30\% and the errors on the derived temperatures are $\sim$10\%.}
\tablenotetext{a}{Strongly dominated by the zodiacal light 
emission, which is $\sim$0.3 Jy at mid-infrared wavelengths.}
\end{deluxetable} 

\clearpage
\begin{deluxetable}{llccclccll}
\rotate
\tablecolumns{10}
\tablecaption{CO $J$=3--2 observations on source \label{co3_2}}
\tablewidth{0pt}
\label{tab_{c}odata}
\tablehead{
\colhead{}& \multicolumn{3}{c}{$^{12}$CO 3--2} & &\multicolumn{3}{c}{$^{13}$CO 3--2}&\colhead{}&\colhead{}\\
\cline{2-4} \cline{6-8} \\
\colhead{Name}& \colhead{$\int T_{\rm{MB}}dV$} & \colhead{$V_{\rm{LSR}}$} 
&\colhead{$\Delta V$} & &\colhead{$\int T_{\rm{MB}}dV$} & \colhead{$V_{\rm{LSR}}$} 
&\colhead{$\Delta V$}&\colhead{$\bar{\tau}$\tablenotemark{a}}&\colhead{$M_{\rm{disk}}$($^{13}$CO)}\\

\colhead{}& \colhead{(K~km s$^{-1}$)} & \colhead{(km s$^{-1}$)} & 
\colhead{(km~s$^{-1}$)} &  &\colhead{(K~km s$^{-1}$)} & \colhead{(km s$^{-1}$)} 
&\colhead{(km~s$^{-1}$)}&\colhead{}&\colhead{(10$^{-3}$ M$_{\odot}$)}}
\startdata
DM~Tau   & \phm{$<\ $}1.02$\pm$0.30                    & 6.2     & 1.0     & & \phm{$<\ $}0.25$\pm$0.07  & 6.4 & 0.6   & 0.56 & 0.18$\pm$0.05 \\
         & \phm{$<\ $}\nodata                          & \nodata & \nodata & & \phm{$<\ $}0.19$\pm$0.06  & 5.6 & 0.6   & \nodata      & \nodata \\
DR~Tau   & \phm{$<\ $}3.45$\pm$1.03                    & 6.8     & 0.7     & & \phm{$<\ $}0.21$\pm$0.06  & 6.9 & 0.3   & 0.06 & 0.07$\pm$0.02\\
         & \phm{$<\ $}1.93$\pm$0.58                    & 9.1     & 0.5     & & \phm{$<\ $} \nodata               & \nodata     & \nodata       & \nodata      & \nodata \\
         & \phm{$<\ $}6.84$\pm$2.05\tablenotemark{c}   & 10.3    & 1.0     & & \phm{$<\ $} \nodata               & \nodata     & \nodata       & \nodata      & \nodata \\
         & \phm{$<\ $}1.38$\pm$0.41\tablenotemark{c}   & 10.0    & 0.9     & & \phm{$<\ $} \nodata               & \nodata     & \nodata       & \nodata      & \nodata \\
GG~Tau   & \phm{$<\ $}1.28$\pm$0.38                    & 5.7     & 1.0     & & \phm{$<\ $}0.21$\pm$0.06  & 5.5 & 1.2   & 0.19 & 0.16$\pm$0.05\\
         & \phm{$<\ $}1.41$\pm$0.42                    & 7.0     & 1.0     & & \phm{$<\ $}0.27$\pm$0.08  & 7.2 & 1.2   & \nodata      & \nodata \\
GO~Tau   & \phm{$<\ $}0.77$\pm$0.23                    & 5.2     & 1.0     & & \phm{$<\ $}0.11$\pm$0.03  & 4.3 & 0.8   & 0.18 & 0.09$\pm$0.03\\  
         & \phm{$<\ $}1.31$\pm$0.39                    & 7.1     & 1.0     & & \phm{$<\ $}0.09$\pm$0.03  & 7.0 & 0.3   & \nodata      &  \nodata  \\ 
         & \phm{$<\ $}0.18$\pm$0.05\tablenotemark{c}   & 6.2     & 0.4     & & \phm{$<\ $} \nodata               & \nodata     & \nodata       & \nodata      & \nodata \\
         & \phm{$<\ $}0.09$\pm$0.03\tablenotemark{c}   & 5.5     & 0.7     & & \phm{$<\ $} \nodata               & \nodata     & \nodata       & \nodata      & \nodata \\ 
RY~Tau   & \phm{$<\ $}3.94$\pm$1.18                    & 6.3     & 0.3     & & \phm{$<\ $}0.21$\pm$0.06  & 6.4 & 0.3  &  \nodata      &0.06$\pm$0.02\\
         & \phm{$<\ $}2.48$\pm$0.74                    & 6.9     & 0.3     & & \phm{$<\ $} \nodata             &  \nodata           &  \nodata              & \nodata \\
GM~Aur   & \phm{$<\ $}0.62$\pm$0.18                    & 4.8     & 1.0     & & \phm{$<\ $}0.24$\pm$0.07  & 4.6 & 1.6   & 0.35 & 0.16$\pm$0.05\\
         & \phm{$<\ $}0.89$\pm$0.27                    & 6.4     & 0.9     & & \phm{$<\ $}0.21$\pm$0.06  & 6.9 & 1.5   & \nodata      & \nodata \\
LkCa~15  & \phm{$<\ $}0.58$\pm$0.17                    & 5.4     & 1.3     & & \phm{$<\ $}0.16$\pm$0.05  & 5.2 & 1.4   & 0.38 & 0.14$\pm$0.04\\
         & \phm{$<\ $}0.61$\pm$0.18                    & 7.0     & 1.3     & & \phm{$<\ $}0.22$\pm$0.06  & 7.1 & 1.4   & \nodata  & \nodata \\
\\
\tableline
\\
HD~163296 & \phm{$<\ $}1.44$\pm$0.43                   & 4.7     & 1.5  & & \phm{$<\ $}0.43$\pm$0.13     & 4.5 & 1.5 & 0.62 & 0.56$\pm$0.16\\
          & \phm{$<\ $}1.63$\pm$0.49                   & 6.9     & 1.5  & & \phm{$<\ $}0.51$\pm$0.15     & 7.3 & 1.5  & \nodata      & \nodata \\
          & \phm{$<\ $}0.75$\pm$0.22 \tablenotemark{c} & 5.5     & 7.0  & & \phm{$<\ $}\nodata  & \nodata &\nodata  & \nodata \\ 
          & \phm{$<\ $}0.83$\pm$0.25 \tablenotemark{c} & 5.5     & 7.0  & & \phm{$<\ $}\nodata  & \nodata & \nodata & \nodata & \nodata \\
CQ~Tau    & $<$ 0.06                                   &\nodata  & \nodata  & & $<$ 0.06\tablenotemark{b} &\nodata   & \nodata    & \nodata  & \nodata \\
MWC~480   & \phm{$<\ $}1.25$\pm$0.37                   & 4.2     & 1.1  & & \phm{$<\ $}0.27$\pm$0.08     & 4.0 & 1.2 & 0.27 & 0.17$\pm$0.05\\
          & \phm{$<\ $}1.12$\pm$0.33                   & 6.0     & 1.1  & & \phm{$<\ $}0.30$\pm$0.09     & 6.2 & 1.2  & \nodata      & \\
HD~36112  & \phm{$<\ $}1.03$\pm$0.31                   & 4.9     & 4.7  & & \phm{$<\ $}0.31$\pm$0.09     & 5.9 & 1.8 & 0.36 & 0.23$\pm$0.07\\
AB~Aur    & \phm{$<$}26.1$\pm$7.8                      & 5.8     & 1.5  & & \phm{$<\ $}5.00$\pm$1.50     & 5.8 & 1.6 & 0.21 & 1.72$\pm$0.51\\
V892~Tau  & \phm{$<\ $}2.18$\pm$0.65                   & 7.0     & 1.1  & & \phm{$<\ $}\nodata & \nodata & \nodata &\nodata  & \nodata \\
          & \phm{$<\ $}2.27$\pm$0.68                   & 8.2     & 1.2  & & \phm{$<\ $}\nodata & \nodata & \nodata &\nodata  & \nodata \\
\\
\tableline
\\
HD~135344 & \phm{$<$}0.39$\pm$0.12                  & 6.4    & 1.0  & & \phm{$<\ $}\nodata  &\nodata  &\nodata &\nodata  & 2.1$\pm$0.6$\times$10$^{-3}$ \tablenotemark{d} \\ 
          & \phm{$<$}0.41$\pm$0.12                  & 7.7    & 1.0  & & \phm{$<\ $}\nodata & \nodata & \nodata &\nodata  & \nodata\\
\enddata
\tablenotetext{a}{Beam-averaged optical depth of $^{13}$CO line from 
$^{12}$CO/$^{13}$CO ratio, 
assuming $T_{\rm{ex}}$($^{12}$CO)=$T_{\rm{ex}}$($^{13}$CO).}
\tablenotetext{b}{No line detected; the 3$\sigma$ upper limit is computed by
 assuming a total line width of 3 km~s$^{-1}$.}
\tablenotetext{c}{Extended cloud emission}
\tablenotetext{d}{Mass computed from $^{12}$CO 3--2 emission.} 
\end{deluxetable} 

\clearpage
\begin{deluxetable}{llll}
\tablecolumns{4} 
\tablewidth{0pt} 
\tablecaption{$^{12}$CO 6--5 line parameters. \label{co6_5}} 
\tablehead{
\colhead{Name}&\colhead{$\int T_{\rm{MB}}dV$}& \colhead{$V_{\rm{LSR}}$}&\colhead{$\Delta V$}\\ 
\colhead{}& \colhead{(K~km~s$^{-1}$)} &
\colhead{(km~s$^{-1}$)} & \colhead{(km~s$^{-1}$)}} 
\startdata 
DL~Tau & $<$1.8 & \nodata & \nodata \\ 
DM~Tau & $<$1.8 & \nodata & \nodata \\
DR~Tau & \phm{$<$}11.6$\pm$1.5 & 10.0 & 1.6 \\ 
GG~Tau &\phm{$<$}1.9$\pm$0.4 & 4.9 & 3.5 \\ 
GO~Tau & \phm{$<$}4.7$\pm$1.3 & 5.2 & 2.4 \\ 
RY~Tau & $<$2.0 & \nodata & \nodata \\ 
GM~Aur & \phm{$<$}2.8$\pm$0.7 & 5.1 & 1.6 \\ 
LkCa~15 & \phm{$<$}1.9$\pm$0.8 & 6.8 & 3.3 \\ 
CQ~Tau & $<$2.8 & \nodata & \nodata \\ 
MWC~480 & \phm{$<$}2.3$\pm$0.8 & 4.9 & 2.5 \\ 
AB~Aur & \phm{$<$}51.7$\pm$2.2 & 5.9 & 2.1 \\ 
V892~Tau & \phm{$<$}11.7$\pm$0.7 & 7.4 & 1.2\\ 
\enddata
\tablecomments{3$\sigma$ upper limits computed assuming a 
line width $\Delta V$=3 km~s$^{-1}$.}
\end{deluxetable} 

\clearpage
\footnotesize
\begin{deluxetable}{llll}
\tablecolumns{6} \tablecaption{Disk mass deduced from 1.3mm flux. \label{mmcont}} 
\tablewidth{350pt} 
\tablehead{\colhead{Name} & \colhead{$M_{\rm{disk}}$\tablenotemark{a} \ (10$^{-2}$M$_{\odot}$)} & \colhead{F$_{1.3mm}$ (mJy)} & \colhead{Ref.}}
\startdata
AA~Tau       & 1.7$\pm$0.8 &  88$\pm$26 &  1 \\
DM~Tau       & 2.1$\pm$0.9    & 109$\pm$33 &  1  \\
DR~Tau       & 3.1$\pm$1.4 & 159$\pm$48 &  1   \\
GG~Tau       & 11.6$\pm$5.2     & 593$\pm$178 &  2  \\
GO~Tau       & 1.6$\pm$0.7     & 83$\pm$25  &  1 \\
RY~Tau       & 4.0$\pm$1.8    & 229$\pm$69 &  1 \\
GM~Aur       & 4.9$\pm$2.2    & 253$\pm$76 &  1  \\
LkCa~15      & 3.3$\pm$1.5    & 167$\pm$50 &  3  \\
TW~Hya       & 1.5$\pm$0.7    & 784$\pm$235 \tablenotemark{b} &  4 \\
\\
\tableline
\\
UX~Ori       & 4.2$\pm$1.9   &  23$\pm$7 &  6  \\
HD~163296    & 6.5$\pm$2.9   & 441$\pm$132 &  7   \\
CQ~Tau       & 2.2$\pm$1.0   & 221$\pm$66 &  7   \\
MWC~480      & 2.2$\pm$1.0   & 131$\pm$39 &  7   \\       
MWC~863      & 1.0$\pm$0.5   &  45$\pm$13 &  7   \\
HD~36112     & 2.9$\pm$1.3   & 72$\pm$21  &  7    \\
AB~Aur       & 2.1$\pm$0.9   & 100$\pm$30 &  8     \\
WW~Vul       & 3.2$\pm$1.4   &  10.5$\pm$3.1 & 9  \\
V892~Tau     & 5.6$\pm$2.5   & 289$\pm$87 &  8   \\
\\
\tableline
\\
49~Ceti           & 0.04$\pm$0.018   & 12.7$\pm$3.8 & 10  \\
HD~135344        & 0.28$\pm$0.126    & 142$\pm$42   & 11  \\
$\beta$ Pictoris & 0.003$\pm$0.00135   &  24$\pm$7    & 12   \\
\enddata
\tablenotetext{a}{Assuming a dust temperature of 30~K except for 
HD~135344 and $\beta$~Pictoris for which the SED are well fitted by a single 
modified blackbody at 95 and 85~K respectively \citep[see ][]{CW95,Dent00}. The errors in the observed fluxes are taken to be 
$\sim$30\% and the errors on the mass are $\sim$45\%}
\tablenotetext{b}{Flux at 1.1 mm}
\tablerefs{
(1) \citealt{BSCG90};
(2) \citealt{Gui99};
(3) \citealt{OB95};
(4) \citealt{WSD89};
(5) \citealt{Hen94};
(6) \citealt{Natta99};
(7) \citealt{MS97};
(8) \citealt{Hen98};
(9) \citealt{Natta97};
(10) \citealt{Bockelee95};
(11) \citealt{SSBM96};
(12) \citealt{CKSTK91}}
\end{deluxetable} 
\normalsize
\clearpage
\begin{deluxetable}{llll}
\tablecolumns{4}
\tablecaption{Stellar ages\label{ages}}
\tablewidth{0pt}
\tablehead{\colhead{Name}&\colhead{Stellar age\tablenotemark{a}}&\colhead{Previous estimate}&\colhead{Ref.}\\
\colhead{}&\colhead{(Myr)}&\colhead{(Myr)}&\colhead{}\\
}
\startdata
AA~Tau                & 2.4   & \phm{$\simeq$}1.2  & 1\\
DM~Tau                & 2.5   &                    & \\
DR~Tau                & 3.8   & \phm{$\simeq$}2.5  & 1 \\         
GG~Tau a              & 1.7   & \phm{$\simeq$}0.82 & 1\\        
GO~Tau                & 3.2   &                    & \\     
RY~Tau                & 7.8   & \phm{$\simeq$}6.5  & 1\\
GM~Aur                & 1.8   & \phm{$\simeq$}1.3  & \\
LkCa~15               & 11.7  & \phm{$\simeq$}8.3  & 1\\
TW~Hya                & 9.3   &  \phm{$\simeq$}15  & 2 \\
\\
\tableline
\\
UX~Ori                & 4.6   & \phm{$\simeq$}2    & 3 \\
HD~163296             & 6.0   & \phm{$\simeq$}5    & 4 \\
CQ~Tau                & 8.9   & \phm{$\simeq$}10   & 4 \\
MWC~480               & 4.6   & \phm{$\simeq$}6    & 4 \\
MWC~863               & 6.0   & \phm{$\simeq$}5    & 4 \\
HD 36112              & 6.0   & \phm{$\simeq$}6    & 4 \\
AB~Aur                & 4.6   & $\simeq$3--5       & 4 \\
WW~Vul                &\nodata & \phm{$\simeq$}\nodata  & \nodata   \\
V892~Tau              &\nodata & \phm{$\simeq$}\nodata  & \nodata   \\
\\
\tableline
\\
49~Ceti                & 7.8   &  \phm{$\simeq$} \nodata & \nodata \\
HD~135344             & 16.7  &   \phm{$\simeq$}\nodata & \nodata \\
$\beta$~Pictoris      & 20    &   $\simeq$20--100     & 5\\
\enddata
\tablenotetext{a}{The stellar ages were derived using the evolutionnary tracks of \citealt{SFB00}.}
\tablerefs{(1) \citealt{SFB99};
(2) \citealt{Webb99};
(3) \citealt{Natta99};
(4) \citealt{MS97};
(5) \citealt{BSS99}
}
\end{deluxetable}
\clearpage  
\begin{deluxetable}{llll}
\tablecolumns{4}
\tablecaption{Summary of disk gas masses deduced by various techniques\label{summary}}
\tablewidth{0pt}
\tablehead{\colhead{Name}&\colhead{$M_{\rm 1.3mm}$(total)}&\colhead{$M_{\rm CO}$(total)}&\colhead{$M_{\rm H_2}$(warm gas)}\\
\colhead{}&\colhead{(10$^{-3}M_{\odot}$)}&\colhead{(10$^{-3}M_{\odot}$)}&\colhead{(10$^{-3}M_{\odot}$)}\\
}
\startdata
AA~Tau                & 17$\pm$8   & \nodata  & 0.2--20 \\
DM~Tau                & 21$\pm$9   & 0.18$\pm$0.05         & \nodata       \\
DR~Tau                & 31$\pm$14   & 0.07$\pm$0.02         & \nodata       \\         
GG~Tau a              & 116$\pm$52  & 0.16$\pm$0.05         & 3.6$\pm$1.8   \\        
GO~Tau                & 16$\pm$7   & 0.09$\pm$0.03         & 6.4$\pm$3.2   \\     
RY~Tau                & 40$\pm$18   & 0.06$\pm$0.02         & \nodata       \\
GM~Aur                & 49$\pm$22   & 0.16$\pm$0.05         & \nodata       \\
LkCa~15               & 33$\pm$15   & 0.14$\pm$0.04         & 8.6$\pm$4.3   \\
\\
\tableline
\\
UX~Ori                & 42$\pm$19   & \nodata  &    9--117   \\
HD~163296             & 65$\pm$29   & 0.56$\pm$0.16         &  0.4$\pm$0.2 \\
CQ~Tau                & 22$\pm$10   & \nodata         &  2.0$\pm$1.0 \\
MWC~480               & 22$\pm$10   & 0.17$\pm$0.05         &  0.7--78.8   \\
MWC~863               & 10$\pm$5   & \nodata         &  1.5$\pm$0.8 \\
HD 36112              & 29$\pm$13   & 0.23$\pm$0.07         &  0.2--18.7   \\
AB~Aur                & 21$\pm$9   & 1.72$\pm$0.51         &  1.3$\pm$0.7 \\
\\
\tableline
\\
49~Ceti               & 0.4$\pm$0.2  & 10$^{-3}$\      & 0.3--2.3\\
HD~135344             & 2.8$\pm$1.3  & 2.1$\pm$0.6$\times$10$^{-3}$ & 6.4$\pm$3.2\\
$\beta$~Pictoris      & 0.03$\pm$0.015 & \nodata                    & 0.17$\pm$0.08\\
\enddata
\tablecomments{See Tables 3, 5 and 7 for details. The details of the derivations are given in sections 5.4 (for $M_{\rm 1.3mm}$), 5.3 (for $M_{\rm CO}$) and 
4.4 (for $M_{\rm H_2}$).}
\end{deluxetable}


\clearpage

\clearpage

\clearpage

\clearpage

\clearpage

\clearpage

\clearpage

\clearpage

\clearpage

\clearpage

\clearpage

\clearpage

\clearpage

\clearpage

\end{document}